# Open charm analysis
# with the ALICE detector
# in pp collisions at LHC

Open charm analyse
met de ALICE detector
in pp botsingen bij LHC

(met een samenvatting in het Nederlands)

**Proefschrift**

ter verkrijging van de graad van doctor aan de Universiteit Utrecht
op gezag van de rector magnificus, prof. dr. J.C. Stoof, ingevolge
het besluit van het college voor promoties in het openbaar te
verdedigen op maandag 30 november 2009 des middags te 4.15 uur

door

**Cristian George Ivan**

geboren op 17 juni 1980 te Olteniţa, Roemenië

Promotor:          Prof. dr. R. Kamermans

Co-promotoren:     Dr. P.G. Kuijer
                   Dr. ir. M. van Leeuwen

# Contents

# Chapter 1

# Theory

## 1.1 Introduction

The quest for the answer to the question 'What is matter made of?' led to the development of elementary particle physics. Its goals are to understand what are the elementary building blocks of nature and how they interact with each other. The currently accepted theory of particle physics is called the Standard Model (SM) and was established in the late seventies. According to the SM the basic components of matter are two types of particles: quarks and leptons, both fermions. These interact with each other via exchange bosons. Leptons interact only weakly ($W^+$, $W^-$, $Z^0$ exchange) and electromagnetically ($\gamma$ exchange) while quarks interact additionally via the strong force (gluon exchange). A classification of the elementary particles is given in Table 1.1

| FERMIONS | | | | | |
|---|---|---|---|---|---|
| | Leptons | | | Quarks | |
| $\nu_e$ (e-neutrino) | $\nu_\mu$ ($\mu$-neutrino) | $\nu_\tau$ ($\tau$-neutrino) | u (up) | c (charm) | t (top) |
| e (electron) | $\mu$ (muon) | $\tau$ (tauon) | d (down) | s (strange) | b (beauty) |
| BOSONS | | | | | |
| **Forces** | | | **Carrier** | | |
| Electromagnetic | | | $\gamma$ | | |
| Weak | | | $W^+, W^-, Z^0$ | | |
| Strong | | | $g$ | | |

Table 1.1: Classification

The particular sector of the Standard Model that deals with the strong interactions is the theory of Quantum Chromodynamics (QCD) in which quarks and gluons are collectively called partons. The dynamics of the interactions are described in a similar way as in the theory of Quantum Electrodynamics (QED). Besides electric charge, quarks carry an additional type of charge which is called colour and appears in three states: $r$ (red), $g$ (green) and $b$ (blue). The exchange particles, gluons, also carry this quantum number which makes them qualitatively different from the exchange particle in QED, the photon, which carries no charge. As a consequence, the coupling 'constant' of QCD becomes increasingly large as the distance between quarks grows. For large momentum transfers (or, equivalently, small distances) the coupling decreases, a property known as 'asymptotic freedom'.

The result of this property of strong interacting matter is that quarks can not be observed freely, instead they are confined inside mesons and baryons (collectively called hadrons). Mesons consist of a quark-antiquark pair (colour and anti-colour) while baryons consist of three quarks which carry three different colours. Following the chromatic analogy, it is said that in nature only colourless objects can be observed.

Although the statement that baryons consist of three quarks is good enough for an intuitive picture, it has been shown that it is not accurate. If the proton is build up by three quarks ($uud$), it is expected that the momentum of the proton is given by the sum of the quark momenta. Experimental data from the Stanford Linear Accelerator (SLAC) [1] show that, on average, only half of the proton's momentum is accounted for by the quarks. The rest is carried by gluons [2]. Gluons can produce quark-antiquark pairs so that at any given moment there is a finite probability that the proton contains an extra pair of quarks $u\overline{u}$, or $d\overline{d}$, or $s\overline{s}$ or even several such pairs. In principle it can also contain even heavier pairs - $c\overline{c}$, $b\overline{b}$, $t\overline{t}$ - but with a by far smaller likelihood because of their larger mass. The quarks which give rise to the hadron quantum numbers are called "valence quarks", while the additional $q\overline{q}$ pairs are called 'sea quarks'.

In high-energy particle collisions (electron-proton scattering for example) it is possible for a virtual photon to couple to one of the sea quarks and produce in the final state a meson containing that specific type of quark.

Because of their confinement property, quarks are difficult to study directly. One can only infer their properties from the behaviour of the hadrons in various contexts. By creating a system in which hadrons are tightly packed together in extremely small volumes the constituent quarks are no more confined inside one single hadron and the interaction between them becomes weak and they can freely move inside the system. This leads to the creation of a new state of matter which is called a Quark Gluon Plasma (QGP) and the transition is expected to occur at a temperature of about 175 MeV and an energy density of 0.7 GeV/$fm^3$ [3]. Such high energy densities can be achieved in the laboratory through collisions of heavy ions. In recent years the Relativistic Heavy Ion Collider (RHIC) [4], in Brookhaven has been a major source of new experimental results concerning phenomena that take place in the QGP [5].

As stated before, heavier quarks also potentially exist inside hadrons. The probability to find a certain type of parton at a specific momentum fraction $x$ and a given momentum transfer $Q^2$ in a proton is given by the parton distribution function (PDF).

In highly energetic hadron collisions (proton-proton and Pb-Pb) the momentum transfers can be large enough to create hadrons which contain heavier quarks. D and B mesons are examples of hadrons which contain a light quark and a heavy quark: $D^0 = (c\overline{u})$, $B^+ = (u\overline{b})$. They are referred to as open charm and open beauty (open heavy flavour). Particles like the $J/\Psi$ which consist of a heavy quark-antiquark pair $(c\overline{c})$ are called heavy quarkonia (or hidden heavy flavour).

## 1.2 Heavy flavour in high-energy heavy ion collisions

The accelerator built at CERN [6], the Large Hadron Collider (LHC) [7], will be able to accelerate heavy-ions up to energies of $\sqrt{s_{NN}} = 5.5$ TeV. The aim of studying heavy-ion collisions is to understand how collective phenomena and macroscopic properties, involving many degrees of freedom, emerge from the microscopic laws of elementary particles.

At LHC energies it is expected that the formed QGP has a lifetime in the order of 10 fm/c. Because of their large masses, heavy quarks are produced in the early stage of the collision in primary partonic interaction with large virtuality $Q$ which means small time scales ($\Delta\tau \sim 1/Q$). The minimum value of $Q$ in the production of a heavy quark-antiquark pair ($Q_{min} = 2m_Q$) implies a space-time scale in the order of $1/(2m_Q) \sim 0.1$ fm/c for charm (0.02 fm/c for beauty), which is much lower than the expected lifetime of the QGP. After the formation of the quark-gluon plasma no production of heavy flavour is expected to take place and therefore the initially produced heavy quarks experience the full collision history and carry information about the crossed medium. This particular aspect of heavy quarks makes them a very well suited probe for studying the properties of the quark-gluon plasma.

In the standard model, the masses of the quarks inside hadrons is given by the coupling to the Higgs field (bare masses) and by the spontaneous breaking of the chiral symmetry in QCD. For the high density medium, the chiral symmetry is expected to be restored and the quark masses should decrease to their bare values. Figure 1.1 shows that the masses of heavier quarks do not change in the medium.

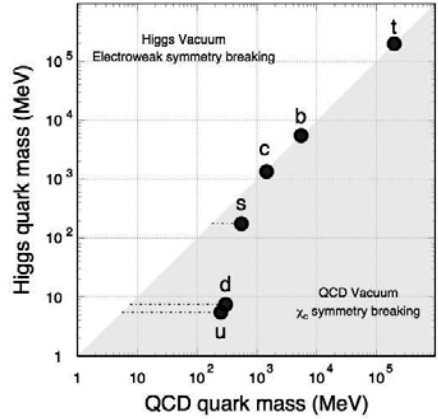

Figure 1.1: Quark masses with contribution from the Higgs field and from QCD symmetry breaking. The mass of heavy quarks is mainly due to Higgs. In the QGP medium, where the chiral symmetry is restored, charm and bottom do not change mass.

In heavy-ion collisions many aspects in the evolution of the system are qualitatively different from proton-proton collisions. The Glauber model [8] of a heavy-ion collision considers the system as a superposition of many *independent* nucleon-nucleon collisions. Thus, the cross section for hard processes in heavy ion collisions can be calculated using a simple geometrical extrapolation from pp collisions, i.e. assuming that the hard cross section scales from pp to nucleusnucleus (AA) collisions proportionally to the number of inelastic nucleonnucleon collisions (binary collisions).

Experimental results show that the QGP behaves qualitatively different from the simple superposition of single hadron collisions. The medium influences the dynamics of all partons involved in the collision and is referred to as nuclear effects. Since heavy flavour is created in the initial stage of the collision and participates to its entire evolution, it is strongly affected by the medium. Therefore departures from binary collision scaling for heavy flavour production in AA provide information about nuclear effects. These can be categorized as: effects due to the embedding of the partons in a nucleus (cold nuclear effects) and effects due to the large energy density in the final state. One focus point for heavy flavour physics in heavy-ion collisions is to investigate the properties of the dense matter by studying its influence on open heavy flavour and quarkonium production.

To quantify the differences between QGP and non-QGP effects, detailed measurements and calculations of $pp$ and $p + A$ heavy flavour cross sections are required. Thus, $pp$ collisions provide important baseline measurements for heavy flavour cross sections and a better understanding of the differences between the mechanisms involved in heavy-ion and nucleon-nucleon collisions.

Dense matter effects in nuclear collisions may change kinematic distributions and total cross section of heavy flavour production. Processes such as energy loss and flow can significantly modify the heavy flavour $p_T$ distribution but do not change the total yields. Energy loss steepens the momentum slope because the heavy quark $p_T$ is reduced. Thus ,in a finite acceptance detector, the measured yields may appear to be enhanced or suppressed, depending on the acceptance. If the momentum of the quarks is small ($p_T < m_Q$) and the surrounding medium exhibits collective motion (such as flow), the quarks may take the velocity of the medium. Measurements at RHIC have shown charm flow in Au+Au collisions [9].

The energy loss mentioned earlier can be due to: elastic collisions with light partons (collisional energy loss) and gluon radiation (radiative energy loss), which is the analogue of bremsstrahlung. Studies suggest that, because of its heavier mass, the bottom quark energy loss should be smaller in the medium than that of the charm quark as Dokshitzer and Kharzeev pointed out that soft gluon radiation from heavy quarks is suppressed at angles $\theta_0 < m_Q/E_Q$ [10]. Thus bremsstrahlung is suppressed for heavy quarks relative to light quarks by the factor $(1 + \theta_0^2/\theta^2)^{-2}$, the so called 'dead cone' effect. Because the energy loss for lighter quarks is stronger than for heavy flavour, the latter can be used to probe deeper into the medium.

Figure 1.2 shows a typical process of two colliding hadrons where low-$x$ partons create a pair of heavy quarks. Due to the high energies used at the LHC one can probe the parton distribution function of the nucleon down to values in the order of $10^{-4}$ for measurements at central rapidities. Because of its lower mass, charm allows one to probe lower values of $x$ than beauty. For an extended rapidity measurement ($|y| \simeq 4$) it is possible to access $x$ regimes about two orders of magnitude lower, down to $x \sim 10^{-6}$ [11]. Measurements at these scales may provide important information about the postulated extreme state of matter described by the theory of the color glass condensate (CGC) [12, 13]. Table 1.2 summarizes the Bjorken-$x$ values corresponding to charm and beauty production at central rapidity and transverse momentum $p_T \rightarrow 0$ for SPS, RHIC and the expected ones for LHC.

In case of $pA$ and $AA$ collisions the modification of PDF's in the nucleus can be determined. The cross section of charm and beauty at LHC will be significantly affected by the parton dynamics in the region of small Bjorken-$x$.

| Machine | **SPS** | **RHIC** | **LHC** | **LHC** |
|---|---|---|---|---|
| System | Pb-Pb | Au-Au | Pb-Pb | pp |
| $\sqrt{s_{NN}}$ | 17 GeV | 200 GeV | 5.5 TeV | 14 TeV |
| $c\bar{c}$ | $x \simeq 10^{-1}$ | $x \simeq 10^{-2}$ | $x \simeq 4 \times 10^{-4}$ | $x \simeq 2 \times 10^{-4}$ |
| $b\bar{b}$ | - | - | $x \simeq 2 \times 10^{-3}$ | $x \simeq 6 \times 10^{-4}$ |

Table 1.2: Bjorken-$x$ values corresponding to charm and beauty production at central rapidity and $p_T \to 0$ at SPS, RHIC and LHC energies [11].

## 1.3 Heavy flavour in proton-proton collisions

In QCD the production of b quarks (and to a lesser extent also c quarks) can be predicted using perturbative QCD, since their masses are large enough to assure the reliability of the calculations ($m_Q > \Lambda_{QCD}$).

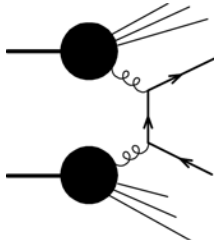

Figure 1.2: Typical diagram of heavy quarks creation.

The measurement of heavy flavour production in *pp* collisions provides valuable information about the mechanisms involved in the parton interaction. Current theoretical models do not give a complete and accurate description of such mechanisms as non-perturbative phenomena have important contribution to heavy flavour production. The available calculations are performed by matching the resummation of logarithms of the transverse momentum over the mass of the quark at next-to-leading-logarithm (NLL) accuracy with the fixed-order, exact NLO calculations for massive quarks (FONLL).

Computer models such as PYTHIA [14] or HERWIG [15] are using perturbative QCD calculations (pQCD) exact only at leading order and include processes like pair creation ($q\bar{q} \to Q\overline{Q}$ and $gg \to Q\overline{Q}$). In PYTHIA, processes above the leading order (NLO, shorthand for next-to-leading-order) are also included and are able to reproduce well some aspects of NLO effects.

Some of the processes taken into account by PYTHIA are illustrated in Fig. 1.3: pair creation, flavour excitation ($qQ \rightarrow qQ$ and $gQ \rightarrow gQ$) and gluon splitting ($g \rightarrow Q\overline{Q}$).

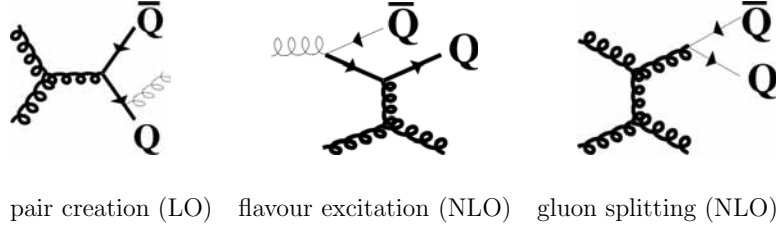

pair creation (LO)    flavour excitation (NLO)    gluon splitting (NLO)

Figure 1.3: Leading order and next-to-leading order processes of heavy-flavour creation included in PYTHIA. Thick lines represent hard processes and thin lines correspond to initial or final state parton shower.

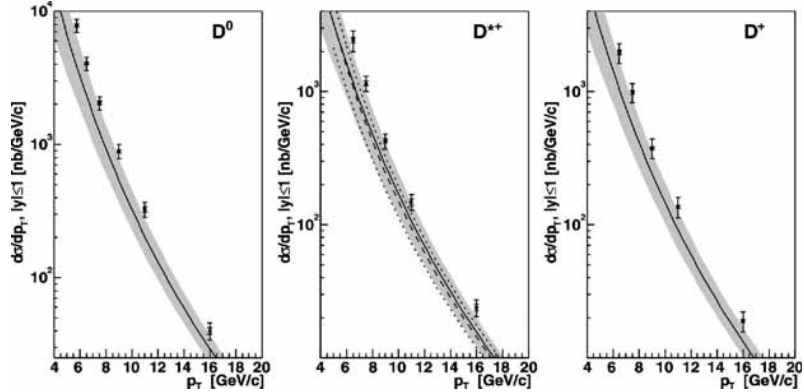

Figure 1.4: Measured differential cross section for $p\overline{p}$ collisions at $|y| \leq 1$ and $\sqrt{s} = 1.96$ TeV [16] at CDF. Solid curves are FONLL predictions by Cacciari and Nason [17] with the uncertainties indicated by the shaded bands.

Heavy flavour measurements have been done by various experiments at Tevatron [16] and RHIC [18]. Figure 1.4 shows the D meson production measurements at $\sqrt{s} = 1.96$ TeV and $|y| \leq 1$ compared to FONLL calculations. The solid curves are the theoretical predictions from Cacciari and Nason [17], with the uncertainties indicated by the shaded bands. The dashed curve shown with the $D^{*+}$ cross section is the theoretical prediction from Kniehl [19] where the dotted lines indicate the uncertainty. The measured differential cross sections are higher than the theoretical predictions by about 100% at low $p_T$ and 50% at high $p_T$. However, they are compatible with the calculations within uncertainty bands.

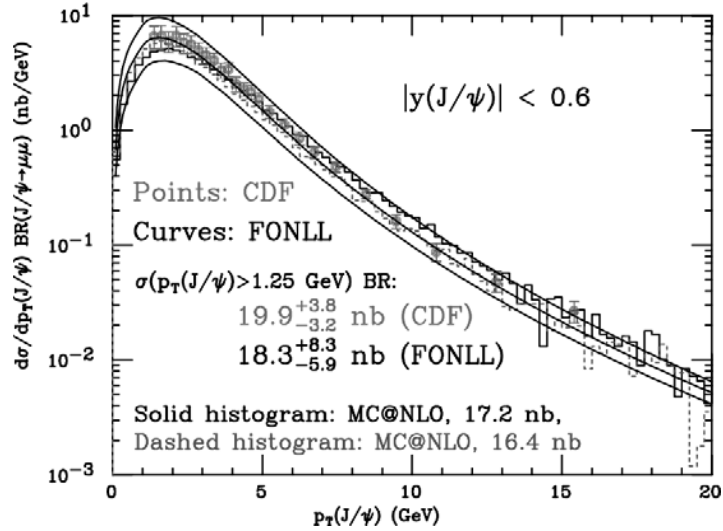

Figure 1.5: Transverse momentum for $J/\Psi$ from B decays as measured by CDF for $p\bar{p}$ at $\sqrt{s} = 1.96$ TeV. Data is compared to theoretical predictions (the band represents FONLL systematic uncertainties) and MC@NLO calculations using two different $b$ hadronization parameters [20].

In contrast to the charm data, measurements for beauty production are better described by theory [20]. Figure 1.5 shows the $J/\Psi$ transverse momentum distribution from inclusive $B \to J/\Psi + X$ decays as measured by the CDF collaboration for rapidity $|y_{J/\Psi}| < 0.6$. The data points lie well within the uncertainty band, and are in very good agreement with the central FONLL prediction. Also shown are two predictions from the MC@NLO code [21, 22] corresponding to two choices of the $b$ hadronization parameters considered in [20]. Despite the good accuracy of partonic calculations, the comparison with data is affected by the presence of non-perturbative ingredients needed to parametrize the fragmentation of heavy quarks into the observed hadrons and by limited phase space accessible to present detectors.

Measurements taken by the PHENIX collaboration [23] for $pp$ collisions at $\sqrt{s} = 200$ GeV are shown in Fig. 1.6 which shows the invariant differential cross section for the production of electrons from heavy-flavour decays.

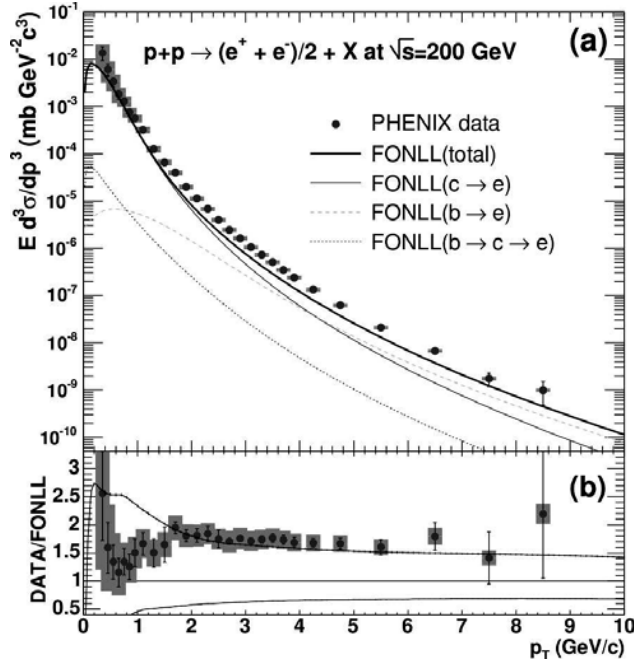

Figure 1.6: **(a)** Invariant differential cross sections of electrons from heavy-flavour decays. The error bars (bands) represent the statistical (systematic) errors. The curves are the FONLL calculations. **(b)** Ratio of the date and the FONLL calculation. The upper (lower) curve shows the theoretical upper (lower) limit of the FONLL calculation.

The curves represent the central values of the FONLL calculations for total charm and beauty [24]. Separate curves for charm electrons, electrons from beauty and the contribution of $b$ to charm electrons $(c \rightarrow e)$, $(b \rightarrow e)$ and $(b \rightarrow c \rightarrow e)$ show the individual contribution of $b$ and $c$ to the total spectrum. The results show that for $p_T > 4$ GeV/c, beauty production becomes dominant.

In Fig. 1.6 (b), the ratio of the data to the FONLL calculation is shown. The ratio shows a slight $p_T$ dependency over the entire range [18]. Fitting with a constant for $0.3 < p_T < 9.0$ GeV/c yields a ratio of ∼1.72. Similar ratios were observed at the Tevatron [16].

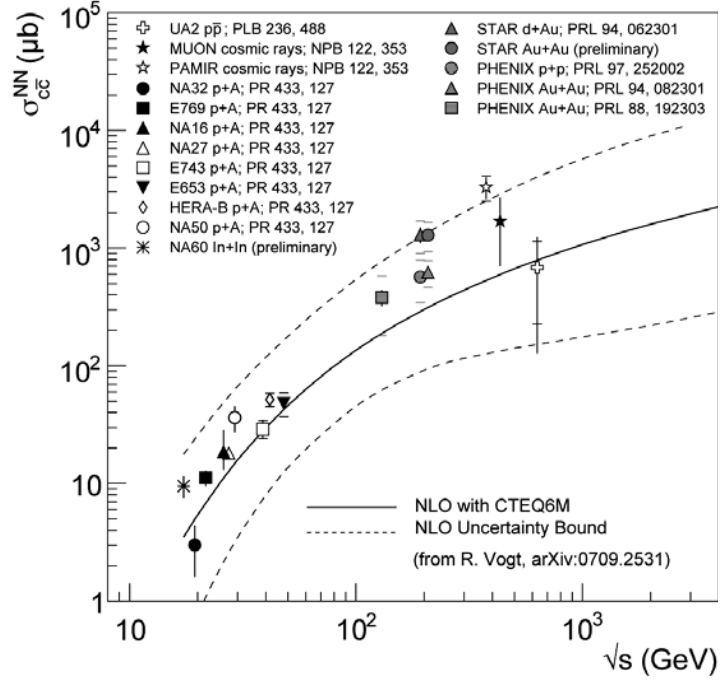

Figure 1.7: Comparison of total charm production cross section measurements [25].

Both PHENIX and STAR [26] have measured charm production cross sections at mid-rapidity. These measurements have been extrapolated to all rapidities to yield total cross sections. These total cross sections are compared to results at other energies and to pQCD calculations in Fig. 1.7 [25]. The results for d+Au and Au+Au collisions are divided by the number of binary collisions, $N_{coll}$, to directly compare with the pp results. The STAR data are from combined fits to hadronic and semileptonic decay data. The PHENIX data are from semileptonic decay measurements only. The STAR values are somewhat higher than those for PHENIX. The total charm cross sections differ by about 1.5 times the combined standard deviations of the two measurements, obtained by adding all statistical and systematic uncertainties in quadrature.

Comparisons between theoretical predictions and experimental results from the Tevatron and RHIC show that the mechanisms involved in the production of heavy flavour are not well understood. In the context of separating cold nuclear effects from QGP effects, a deeper understanding of heavy flavour production in nucleon-nucleon collisions is needed.

Additional effects expected to occur at low-$x$ may shed some light on the phenomena occurring in non-perturbative regimes which were not experimentally reachable in previous experiments.

Differences between heavy-ion and nucleon-nucleon collisions are seen at various levels and affect many observables. **Particle multiplicities** in $pp$ and $AA$ are related to differences between parton distributions in the nucleon with respect to the nuclear case (shadowing effects) in part due to saturation of parton densities at small-$x$. Models predict that the presence of in-medium parton energy loss affects jets produced in $AA$ with respect to pp. It is expected that jets in $AA$ are heavily suppressed compared to the case of $pp$, phenomenon called **jet quenching**.

Heavy flavour production in proton-proton and proton-nucleus collisions, besides proving a benchmark study for medium effects, are important as a test to both perturbative and non-perturbative sectors of QCD in a new energy domain.

## 1.4 D mesons

The production of open charm is investigated by reconstructing $D^0$ and $D^{*+}$ mesons. B meson decays also contribute to the production yield of D mesons. A more detailed discussion about the ratio $(D^0$ from b$)/(D^0$ from c$)$ is given in chapter 5.

Relevant channels for the $D^0$ production from charm and beauty are:

$$c \to D^0 + X \qquad b \to B \to D^0 + X$$
$$c \to D^{*+} + X \qquad b \to B \to D^{*+} + X \qquad (1.1)$$

We used PYTHIA as a Monte-Carlo event generator to study the relevant decay channels. The way in which D mesons are treated will be described in the following. PYTHIA assumes that charm quarks fragment to D (spin singlets: $J = 0$) and $D^*$ (spin triplets: $J = 1$) mesons according to the number of available spin states: $N_{D^0}:N_{D^+}:N_{D^{*0}}:N_{D^{*+}} = 1{:}1{:}3{:}3$. The difference between neutral and charged D mesons arises here: owing to the slightly larger ($\simeq 4$ MeV/c$^2$) mass of the $D^+$, the $D^{*+}$ decays preferably to $D^0$ and the $D^{*0}$ decays exclusively to $D^0$. The total $D^0$ yield is given by the prompt $D^0$ production and decays of $D^{*+}$ and $D^{*0}$. Based on the above considerations, the ratio of neutral to charged mesons is:

$$\frac{D^0}{D^+} = \frac{D^0_{primary} + D^{*+}_{\to D^0} + D^{*0}_{\to D^0}}{D^+_{primary} + D^{*+}_{\to D^+} + D^{*0}_{\to D^+}} = 3.08 \ . \tag{1.2}$$

Following the same reasoning the ratio of $D^0$ to $D^{*+}$ is 2.01. The yields for the $D^0$ and $D^+$ mesons are listed in table 1.3.

The ALEPH Collaboration [27] found that for $e^+e^-$ collisions the ratio $D^0/D^+$ is $\simeq 2.4$, lower than the expected value. This shows that naïve spin counting does not hold. Assuming that the same effect would be present at LHC, we would expect the total $D^0$ yield to diminish by about 6%:

$$\left(\frac{D^0}{D^0+D^+}\right)^{PYTHIA} \Big/ \left(\frac{D^0}{D^0+D^+}\right)^{ALEPH} = 0.94 \tag{1.3}$$

| Particle | Yield | $\langle dN/dy\rangle_{|y_{lab}|<1}$ | Relative abundance |
|---|---|---|---|
| $D^0 + \overline{D^0}$ | 0.1908 | 0.0196 | 61% |
| $D^+ + D^-$ | 0.0587 | 0.0058 | 19% |
| $D_s^+ + D_s^-$ | 0.0362 | 0.0038 | 12% |
| $\Lambda_c^+ + \Lambda_c^-$ | 0.0223 | 0.0026 | 8% |

Table 1.3: PYTHIA calculations for the total yield, average rapidity density for $|y| < 1$ and relative abundance for hadrons with charm in pp collisions at $\sqrt{s} = 14$ TeV.

The rapidity and pseudorapidity of a particle are defined in terms of its energy-momentum components by

$$y = \frac{1}{2}\ln\left(\frac{p_0 + p_z}{p_0 - p_z}\right)$$

$$\eta = \frac{1}{2}\ln\left(\frac{|\mathbf{p}| + p_z}{|\mathbf{p}| - p_z}\right) \tag{1.4}$$

and in the limit of large momentum they coincide. Measurements are done for rapidity $|y| < 1$ because in colliders most of the produced particles are at mid-rapidity.

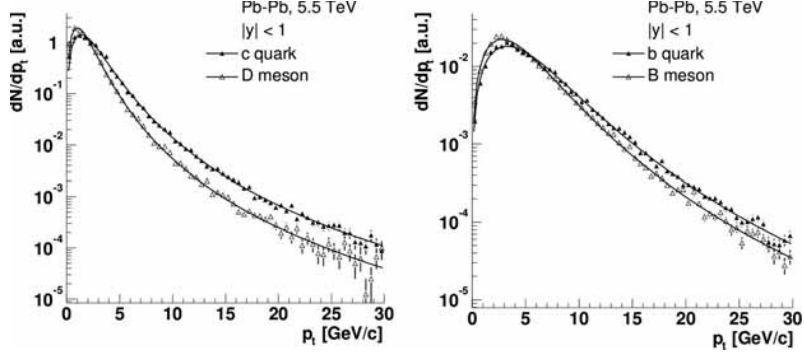

Figure 1.8: Transverse momentum distributions at mid-rapidity for heavy quarks and mesons in Pb-Pb at $\sqrt{s_{NN}} = 5.5$ TeV. Histograms are normalized to the same integral.

Figure 1.8 shows the transverse momentum distributions for rapidity ($|y| < 1$) for c quarks and D mesons (left) and for b quarks and B mesons (right) in Pb-Pb at $\sqrt{s_{NN}} = 5.5$ TeV. For $p_T > 0$ and $|y| < 1$, we have, on average $p_T^D \simeq 0.75\ p_T^c$ and $p_T^B \simeq 0.75\ p_T^b$. The expression in Eq. 1.5 is fitted to the transverse momentum distribution for the D and B meson

$$\frac{1}{p_T}\frac{dN}{dp_T} \sim \left[1 + \left(\frac{p_T}{p_T^0}\right)^2\right]^{-n} \tag{1.5}$$

with $n$ and $p_T^0$ specified in Table 1.4.

| Particle | System | $\sqrt{s_{NN}}$ (TeV) | $p_T^0$ (GeV/c) | $n$ | $<p_T>$ GeV/c |
|---|---|---|---|---|---|
| D ($|y| < 1$) | pp | 14 | 2.04 | 2.65 | 1.85 |
| | pPb | 8.8 | 2.09 | 2.72 | 1.83 |
| | Pb-Pb | 5.5 | 2.12 | 2.78 | 1.81 |
| B ($|y| < 1$) | pp | 14 | 6.04 | 2.88 | 4.90 |
| | pPb | 8.8 | 6.08 | 2.90 | 4.89 |
| | Pb-Pb | 5.5 | 6.14 | 2.93 | 4.89 |

Table 1.4: Parameters derived from the fit of the $p_T$ distributions of D and B mesons to the expression 1.5 and their average transverse momentum.

# Chapter 2

# The ALICE Detector

**A L**arge **I**on **C**ollider **E**xperiment [28] has been built at CERN to investigate the physics of strongly interacting matter at very high energy densities. The **L**arge **H**adron **C**ollider will accelerate and collide Pb nuclei at an energy of $\sim 5.5$ TeV/nucleon.

The detector is designed to perform tracking and vertexing in an extremely high particle density environment: $dN_{ch}/dy$ up to 8000 [11]. This is considered to be an overestimation of the multiplicity in Pb-Pb collisions at LHC. Before the RHIC measurements at $\sqrt{s_{NN}} = 200$ GeV, models predicted a multiplicity which was a factor two higher compared to what was actually observed: $dN_{ch}/dy \simeq 650$. In view of these results it is now expected that the multiplicity at the LHC will not exceed 4000 particles per unit of rapidity. Fig. 2.1 shows the result which reproduces well the multiplicity measurements done by RHIC. It predicts that the LHC will produce $\simeq 2500$ charged particles per unit of rapidity in central Pb-Pb collisions.

Besides heavy-ion collisions, different systems will be used at various beam energies: p-p, p-A and A-A collisions will be used to set reference data for Pb-Pb collisions. The LHC running programme foresees [29]:

- regular pp runs at $\sqrt{s} = 14$ TeV

- 1-2 years of Pb-Pb at $\sqrt{s} = 5.5$ TeV

- 1 year of p-Pb at $\sqrt{s} = 8.8$ TeV

- 1-2 years of Ar-Ar at $\sqrt{s} = 6.3$ TeV

The ALICE detector consists of a central barrel embedded in a large solenoidal magnet and a muon arm with a separate dipole magnet. The acceptance of the central detector system covers $-0.9 < \eta < 0.9$ over the full azimuth and the muon arm $2.4 < \eta < 4$. The global reference frame is as follows: the $z$ axis is parallel to the beam and pointing to the muon arm, $x$ and $y$ in the transverse plane to the beam direction. The central detector consists of:

- an Inner Tracking System (ITS) with six layer of high-resolution silicon detectors;

- a Time Projection Chamber (TPC);

- a Transition Radiation Detector (TRD) for electron identification;

- a Time-Of-Flight barrel (TOF);

- a single-arm imaging Cherenkov detector for high momentum particle identification (HMPID);

- a single arm electromagnetic calorimeter (PHOS) consisting of high-density crystals;

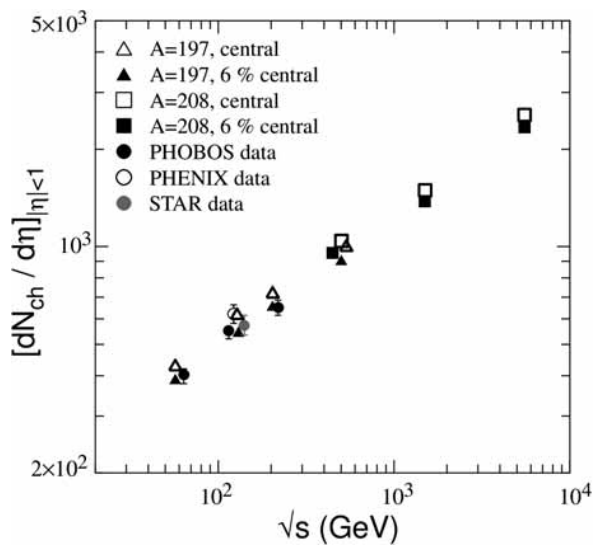

Figure 2.1: Charged multiplicity per unit of pseudorapidity in A-A collisions. Square and triangle markers are model predictions. Circle markers are measurements at RHIC.

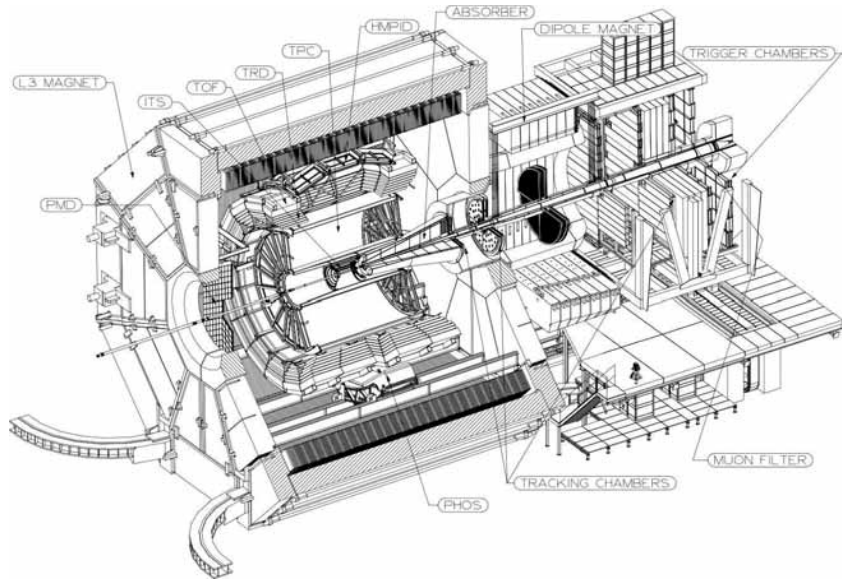

Figure 2.2: A sketch of the ALICE detector.

Several smaller detectors (ZDC, PMD, FMD, T0, V0) are located at forward angles and are used for triggering and general event characterization. The layout of all sub-detectors is shown in Fig. 2.2.

The ALICE magnet provides a magnetic field of 0.5 T, parallel to the beam axis. The field has been chosen to have both a good acceptance for low momentum tracks and a good momentum resolution for high $p_T$. A better choice for low momentum particles would be a field of 0.2 T, but since statistics at high momentum is poor, the experiment will run with 0.5 T.

The beam pipe is made from beryllium, a low atomic number element, to minimize multiple scattering in the material, and has an outer radius of 3 cm and a thickness of 0.8 mm.

The ALICE detector is designed to detect particles in a broad range of transverse momentum (100 MeV/c $< P_t <$ 100 GeV/c) and reconstruct the primary vertex position with an accuracy better than 100 $\mu m$. The study of heavy flavour ($b$ and $c$ quarks) requires a high resolution vertex detector close to the beam pipe.

## 2.1   Inner Tracking System (ITS)

The main goals for which the ITS was designed are:

- primary and secondary vertex reconstructions for the detection of neutral particles, cascade hyperon decays, charm and beauty meson decays,

- tracking in low transverse momentum region ($70 < P_t < 200$ MeV/c) as a stand-alone tracker,

- particle-identification with $dE/dx$ measurements,

- improvement in momentum resolution for high $P_t$ particles.

This is achieved by using a 6 layer detector system. The positioning of the 6 concentric layers is optimized for tracking and spatial resolution at various distances from the primary vertex (Fig. 2.3)

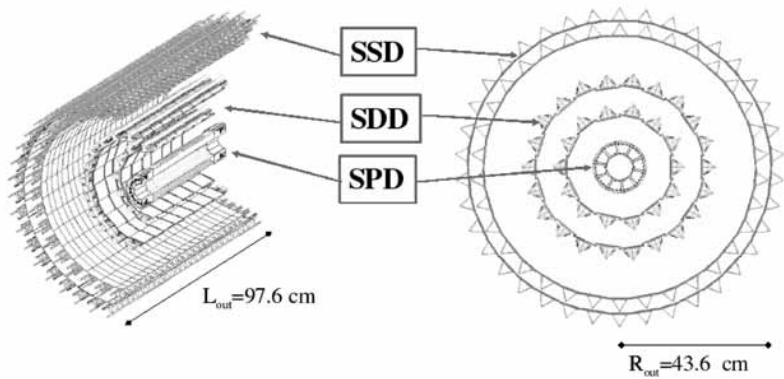

Figure 2.3: Axonometric and front-side view of the ITS. Spatial arrangement of the layers can be seen.

The first two layers are positioned at 4 and 7.2 cm from the interaction point. The high particle density in this region ($> 50$ tracks/cm$^2$) requires an excellent position resolution which is achieved with a silicon pixel detector (SPD) with the cell size of $50(r\phi) \times 425(z)$ $\mu$m$^2$.

At r=15 and 23.9 cm there are two layers of silicon drift detectors (SDD) and at r=38.5 and 43.6 cm two layers of silicon strip detectors (SSD). The SSD is capable of delivering good tracking together with energy loss measurements. At larger distances form the interaction point (where the particle density can be lower than one track per cm$^2$) the double-sided microstrip detectors also allow for $dE/dx$ measurements and delivers important information for the connection of tracks from the TPC and ITS.

The acceptance of the ITS is of $|\eta| < 0.9$ for all vertices located in the region of the interaction diamond (-5.3 < z < 5.3 cm along the beam direction). The first layer of pixel detectors have a larger range of rapidity coverage. Together with the forward multiplicity detectors it allows for a continuous measurement of charge multiplicity in the range -3.4 < $\eta$ < 5.1.

| Layer | Type | r(cm) | ±z(cm) | $|\eta|$ | $\sigma_{r\phi}(\mu m)$ | $\sigma_z(\mu m)$ |
|-------|------|-------|--------|----------|-------------------------|-------------------|
| 1 | Pixel | 4.0 | 14.1 | 1.98 | 12 | 100 |
| 2 | Pixel | 7.2 | 14.1 | 0.9 | | |
| 3 | Drift | 15.0 | 22.2 | 0.9 | 38 | 28 |
| 4 | Drift | 23.9 | 29.7 | 0.9 | | |
| 5 | Strip | 38.5 | 43.2 | 0.9 | 20 | 830 |
| 6 | Strip | 43.6 | 48.9 | 0.9 | | |

Table 2.1: ITS main characteristics.

## 2.2 Time Projection Chamber (TPC)

The main tracking device of the ALICE central barrel is the TPC. It is an adaptation and improvement of the devices already used successfully in other experiments: STAR, ALEPH, NA49 [26, 27, 30]. It is designed to provide charged-particle momentum measurements with good two-track separation, particle identification, and vertex determination. The global acceptance is $-0.9 < \eta < 0.9$ for full radial tracks and $2\pi$ azimuthal coverage. For tracks with reduced track length and momentum resolution the acceptance can be extended up to $|\eta| \sim 1.5$.

The TPC is required to cope with the Pb-Pb design luminosity, corresponding to an interaction rate of 8 kHz. About 10% of these collisions are considered central. The multiplicity in this case was safely overestimated at 8.000 particles per unit of rapidity, resulting in 20.000 charged primary and secondary tracks in acceptance [11]. This is an unprecedented track density for a TPC.

Properties of the QGP are obtained through various hadronic measurements which include spectroscopy of strange and multi-strange hadrons, single- and two-particle correlations. These impose high demands on $p_T$ acceptance, rapidity, particle identification and azimuthal coverage. Good momentum resolution at high-$p_t$ is required by the analysis of hard probes (heavy quarkonia, charmed, beauty particles and high-$p_T$ jets). The momentum resolution can be achieved by combining the TPC with other tracking detectors.

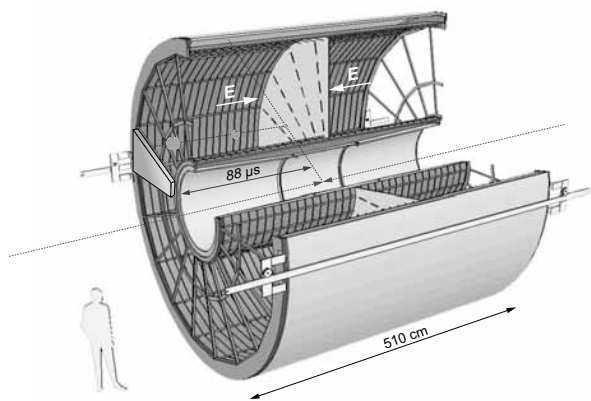

Figure 2.4: Layout of the ALICE TPC.

The aimed perfromance of the TPC for hadronic measurements include:

- *Momentum resolution:* soft physics observables require a resolution of 1% for low momenta ($0.1 < p_T < 1$ GeV/c). Depending on the magnetic field the resolution for low-$p_T$ is between 1 and 2%. For the high-$p_T$ region (required by hard probes) 10% of momentum resolution (as high as 100 GeV/c) can be achieved by using the TPC in conjunction with the ITS and TRD (at 0.5 T magnetic field).

- *Two-track resolution:* Two-particle correlations require a good resolution on the momentum difference measurement: around 5 MeV/c or better for a source size of 20 fm.

- *dE/dx resolution:* For the low momentum region particle identification via *dE/dx* is done in specific momentum intervals where the ionization for different particle types is well separated. For measurements up to a few GeV/c one has to used information from other detectors as well (like the time of flight detector). The resolution of the ionization measurements for higher $p_t$ depends on the particle density and, based on simulations, a value of 6.9% can be achieved for extreme multiplicities.

- *track matching:* For the measurement of the impact parameters at the interaction point and the secondary vertices track matching between the TPC and ITS needs to be very efficient. A good matching with other detectors will help improving the momentum resolution of the TPC-stand-alone reconstruction (up to a factor 5 improvement is expected for 10 GeV/c tracks).

- *Azimuthal coverage:* Flow analyses as well as events and signals with limited statistics require a full azimuthal coverage.

The layout of the detector is shown in Fig. 2.4 and the main characteristics summarized in Table 2.2.

| | |
|---|---|
| Pseudo-rapidity range | $-0.9 < \eta < 0.9$ for full radial track length |
| | $-1.5 < \eta < 1.5$ for 1/3 radial track length |
| Azimuthal coverage | $2\pi$ |
| Radial position (active volume) | $845 < r < 2.466$ mm |
| Length | 5.000 mm |
| Segmentation in $\phi$ | 18 sectors |
| Segmentation in $r$ | 2 chambers/sector |
| Segmentation in $z$ | Central membrane, readout on 2 end-plates |
| Material budget | $X/X_0$=2.5 to 5% for $-0.9 < \eta < 0.9$ |
| Detector gas | 88 $m^3$ of $Ne/CO_2$ (9:1) |
| Drift length | 2 ×2.500 mm |
| Drift field | 400 V/cm |
| Drift velocity | 2.84 cm/$\mu$s |
| Maximum drift time | 88 $\mu$s |
| Event size (for $dN/dy$=8000) | ~60 MB |
| Event size (for pp) | ~1-2 MB (depending on pile-up) |
| Data rate limit | 400 Hz (Pb-Pb minimum bias) |
| Trigger rate limits | 200 Hz (Pb-Pb central events) |
| | 1.000 Hz (p-p events) |
| Position resolution | |
| in $r\phi$ | 1100-800 $\mu$m (inner/outer radii) |
| in $z$ | 1250-1100 $\mu$m |
| $dE/dx$ resolution: | |
| Isolated tracks | 5.5% |
| $dN/dy$=8.000 | 6.9% |

Table 2.2: TPC main characteristics.

## 2.3   Off-Line Framework

Since by the time when this thesis was finalised the ALICE experiment has not started yet, the analysis is entirely based on simulated data. It is important to briefly introduce the architecture of the analysis framework in order to explain how the flow of information is, to our best knowledge, simulating the real experiment. The following sections will give a short description of the ROOT and AliRoot [31] software and some event generators which are the basis from where the entire analysis starts.

### 2.3.1   Root

The development of the ROOT software started in 1994 in the context of the heavy ion experiment NA49 at CERN. It is a platform independent physics analysis application written entirely in the C++ programming language and follows the Object Oriented programming paradigm. This approach allows easy extensions to include other capabilities: remote and distributed analysis, dynamical extensions, user implemented macros and libraries via the built-in C++ interpreter (CINT [32]) etc. It provides several ways of user interaction: graphic interface, command line or batch scripts.

### 2.3.2   AliRoot

AliRoot is the name of the ALICE Off-line framework. Its fundamental role is to reconstruct and analyse data coming from simulations as well as from real interactions. It is capable to fully simulate the real experiment, from the initial collision to the data that will be available to the user.

Each of the main steps in the ALICE experiment in the real world has a correspondent in the virtual world of AliRoot via some external libraries:

- **particle collisions** are simulated by Monte-Carlo **event generators**: PYTHIA, HIJING, GeVSim [14, 33, 34] etc. The simulation produces a list of particles in the final state, with specific momenta and rapidity distributions, propagating away from the interaction point.

- As particles emitted from the primary vertex propagate through the detector they interact with the surrounding material (sensitive areas, cables, cooling pipes, support structures etc.). The **interactions with the detector** are simulated by the **transport code** (Geant). Particle decays, pair creations, ionizations, multiple scattering in the material, energy depositions are all taken into account at this level. Geant not only propagates all particles created during the expansion of the fire-ball, but also new particles created by the interaction with the material. Every particle is followed until it exits the volume of the detector or it reaches a level of energy lower than a certain threshold.

- The first two steps in the simulation process are done via the two external libraries: event generators and transport code. After these steps the ALICE specific code comes into play. The energy deposited in the sensitive areas of the detector is transformed into a **detector response** (hit). Detector misalignments are also taken into account at this step.

- The detector response is then formatted according to the output of the front-end electronics and the data acquisition system. This process is done by the **digitization software** and is also part of the AliRoot specific libraries and subroutines. Electronic noise is also simulated by introducing smearing into the signal. The output of this process is very close to the real data that will be transmitted by the real detector.

- The last step of the analysis chain is common to both simulation and real experiment: the **event reconstruction**. The algorithm reconstructs space points and fits the reconstructed space points and creates track candidates, calculates additional track information (fit parameters, energy loss, particle identification), calculates general event properties (multiplicity, centrality class, primary vertex etc.). These tasks are performed by the AliRoot software which then stores the information in a specific file format (**ALI**CE **E**vent **S**ummary **D**ata - AliESD).

The simulation chain can be executed by invoking a few macros. It can be executed on a local machine, a remote machine or cluster of machines with specific software enabled - GRID, CAF (**C**ern **A**nalysis **F**acility) etc. The output of the simulations can be stored locally or on remote storage elements. Input files for later analysis can be retrieved from the local machine or through the network interface. The amount of computing time for the whole simulation process can take from a few minutes in case of p-p collisions up to a few hours for Pb-Pb depending also on the multiplicity and the number of detectors that are included.

## 2.4   Track reconstruction in ITS-TPC

In the following we will be addressing the part of the simulation process that concerns tracking and vertexing.

Due to the high particle multiplicity expected at LHC, the ALICE tracking procedure is a very challenging task.

The track finding procedure developed for the central barrel is based on a Kalman filter approach which is widely used in high-energy physics experiments. It is a powerful method for simultaneous track recognition and reconstruction. Since it is a local method, at any given point along the track it provides the optimal estimate of the track geometrical parameters at that point. Energy loss and multiple scattering in the material are implemented in a direct and simple way. Also track matching between two detectors benefits from the use of this method (for example between the TPC and ITS or TPC and TRD).

The reconstruction consists of an iterative algorithm: **inward** track finding in the TPC (from the outer part to the inner part, close to the ITS), **matching** to the ITS outer layer and track finding down to the innermost pixel layer, back-propagation and **refitting** in the ITS and to the outer layer of the TPC, **matching** to the TRD and **outward** track finding in the TRD and finally **matching** in the TOF for particle identification. For the present work only TPC and ITS information was used.

Particle trajectories are approximated with helices which are fully described by a vector of 5 parameters: two describe the track geometry in the $z$ direction (along the beam pipe and pointing towards the muon arm) and three describing the geometry in the $xy$ plane (referred to as the *bending plane*). To the track description is added a $(5 \times 5)$ covariance matrix which contains, at any given point, the best estimate of the errors on the track parameters and of their correlations.

The track finding in the TPC starts from the outermost pad row, where the track density is lowest. Seed finding can be calculated with and without primary vertex constraint. It begins with a search for pairs of points at a pad row $i$ and at a pad row $j$ closer to the interaction point (currently $i - j = 20$). For each pair, using the two points and the primary vertex position, the first estimate of the state vector at the outer pad row is calculated. Having the track parametrized, intersection points between the helix and the next 20 pads are searched for. If at least half of the points are found, the candidate is saved as a seed. Then a second seed is found using another pair of rows. The Kalman filter starts first by considering tracks with low curvature (higher $p_T$). These are easier to find because the effect of multiple scattering is inversely proportional to the track momentum. The procedure continues by iterating the following three steps:

- *Prediction:* having the state vector $j-1$ of a track at a certain layer, a prediction of the state $j$ is made by propagating the track to the next layer. The track curvature is modified at this step to account for energy loss and multiple scattering. The covariance matrix is also updated.

- *Filtering:* after the extrapolation to the state j, all clusters with the coordinates inside a search window are considered. For each cluster the vector state is updated and the cluster for which a minimum $\chi^2$-increment was obtained is added to the track (provided that the increment is lower than a given $\chi^2_{max}$).

- *Update:* after the cluster assignment to the track, the state is recalculated as well as the covariance matrix.

Tracks found in the TPC are used as seeds for the Kalman filter applied in the ITS. All clusters from the outermost ITS layer which are in the fiducial road are considered. For each of them a new candidate track is defined and propagated to the next layer, without applying any filtering. In this way, a track-tree with several candidates is built from a single TPC track, and only when the inner pixel layer is reached the filtering is applied and the candidate with the lowest $\chi^2$ per assigned cluster is selected. The assigned clusters are removed and the next TPC track is considered.

### 2.4.1 Primary Vertex

The reconstruction of the primary vertex position is done by using the information from the first two layers of the ITS - the Silicon Pixel Detector.

The interaction diamond at the LHC is parametrized as a Gaussian along the $z$ axis with a $\sigma_z = 5.3$ cm. In the $xy$ plane $\sigma_{xy}$ can vary from $\simeq 15\ \mu m$ up to 75 $\mu m$, depending on the beam parameter $\beta^*$, which also affects the beam luminosity and lifetime.

The primary vertex algorithm uses the $z$ coordinates of the reconstructed space points in the first pixel layer. At the vertex $z$ coordinate $z_{true} = 0$ the distribution is symmetric and its centroid ($z_{cen}$) is very close to the nominal vertex position. If the primary vertex is displaced along the $z$ axis then the correlation between the points in the first and second layer of the SPD is considered (the centroids of the two distributions).

AliRoot simulations showed that the primary vertex resolution depends on the vertex location (in case it is displaced compared to the nominal vertex at the LHC), the particle density and the magnetic field. Dead pixels in the detector do not seem to affect the resolution too much, even if the percentage is as high as 10% (see ref. [11]). In case of Pb-Pb collisions with this method $\sigma_z$ can be up to about 10 $\mu$m and $\sigma_{xy} = 25$ $\mu$m.

For p-p events, due to low multiplicity, the reconstruction of the primary vertex has to be done using 3D track information and thus it can be achieved only after track finding. An extensive analysis is given in [11]. Summarized in Fig. 2.5 are the resolution and reconstruction efficiency of the primary vertex as a function of the event multiplicity.

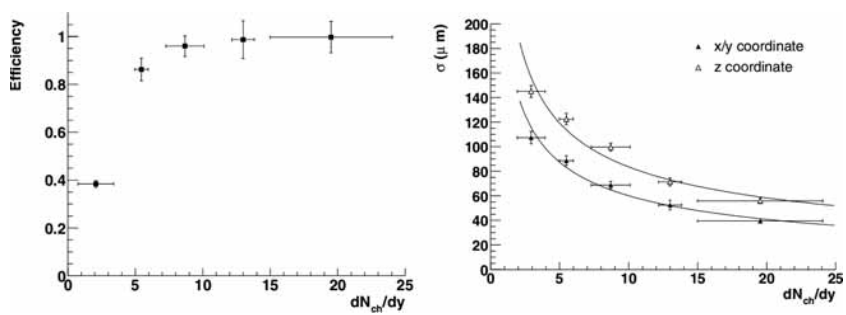

Figure 2.5: Efficiency and resolution for the primary vertex reconstructed using 3D information from reconstructed tracks as a function of the multiplicity in *pp* events.

# Chapter 3

# D meson reconstruction

This chapter starts with a brief discussion about the $D^0$ reconstruction strategy and decay topology. It is then followed by a detailed description of the decay observables and cut optimization procedures. The optimization of cuts for the statistical significance of the $D^0$ and $D^{*+}$ reconstruction, as well as detection efficiencies are described at the end of this chapter.

$D^0$ mesons are reconstructed via the "golden" hadronic channel: $D^0 \rightarrow K^- + \pi^+$ with a branching fraction of 3.8%. The reconstruction is particularly challenging in the case of open charm mesons due to their short lifetimes and therefore very short decay lengths. In the case of the $D^0$ meson the lifetime is around $4.1 \cdot 10^{-13}$ s which gives a $c\tau$ of 123 $\mu m$. Concentrating on particular aspects of the $D^0$ decay one can use the topological properties of the process in order to make a proper reconstruction and analysis. The study of the decay shows various correlations between the topological and kinematic observables. These are used to define selection criteria for the $D^0$ reconstruction.

Apart from the direct $D^0$ production also the production via the $D^{*+}$ has been studied using the decay $D^{*+} \rightarrow D^0 + \pi^+$ with a branching fraction of 68%. This particular process has advantages due to its kinematic properties and will give a good cross-check and an additional benchmark test of the stand-alone $D^0$ analysis. The same properties will be studied as in the case of directly produced $D^0$ mesons.

## 3.1   $D^0$ algorithm and variable description

The $D^0$ reconstruction procedure consists of an event-by-event analysis. All tracks belonging to a single event are combined into oppositely charged pairs in order to reconstruct a possible decay. The loop over the particles starts with the negative tracks ($K^-$) which are then paired with all the other positive tracks ($\pi^+$). The procedure also makes unwanted combinations - it pairs together two tracks which do not represent a real decay and therefore produces a large combinatorial background. One method of reducing the combinatorial background - and hence the CPU time needed to analyse a given set of data - is to impose selection criteria for the tracks. This will ensure that the tracks have reasonable good reconstruction parameters and are good candidates for $D^0$ decay products. These criteria will be described in course of this chapter.

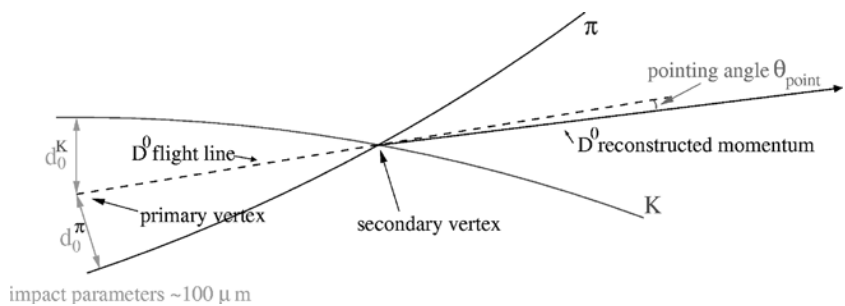

Figure 3.1: Schematic representation of the $D^0$ decay

Figure 3.1 shows a schematic view of the decay process together with some parameters which are directly measured by the detector or can easily be calculated from basic measurements. All notations in the figure will be described in the following.

The "primary vertex" denotes the position where the actual interaction takes place inside the collider's beam pipe and where the charm quark is produced at the time of the collision. This is also the point where the $D^0$ is created. The primary vertex is a key parameter which characterizes the entire event and is extensively used in determining various parameters of the decay.

The decay point of the $D^0$ meson into a ($K^-\pi^+$) pair is referred to as the "secondary vertex". From the primary and secondary vertex one can reconstruct the flight line of the $D^0$. By combining the $K^-$ and $\pi^+$ one can reconstruct the total momentum of the decaying particle. Since the $D^0$ momentum measurement and the flight line are independent, these two vectors will not have the same orientation. The angle between the two is called "pointing angle".

From the helix described by the $K^-$ and $\pi^+$ in the detector's magnetic field one can make a prolongation of the track back towards the primary vertex. The smallest distance from the primary vertex to this extrapolated part of the track in the transverse plane is called the "impact parameter", denoted by $d_0^K$ and $d_0^\pi$ for $K^-$ and $\pi^+$, respectively.

The way the reconstruction of the decay is implemented is based on the above description but with a slight difference in the time order. First, the primary vertex is calculated. Then, two tracks are selected from the event (a negative and a positive one) in order to determine the secondary vertex. The intersection point can not be determined exactly due to two reasons: the selected tracks come from two different decays and are not correlated at all, or they are two decay products of the $D^0$ but due to the reconstruction resolution there is no perfect intersection point defined. As a result, to estimate where the tracks intersect, the distance of closest approach (*dca*) between the two is calculated. The crossing point (secondary vertex) is assumed to lie on the line describing the distance of closest approach.

Another parameter characterizing the decay is the angle between the flight-line of the $D^0$ in the lab frame and the momentum of the $\pi^+$, after the decay, in the rest frame of the $D^0$. Figure 3.4 (left panel) shows a schematic representation of the decay angle ($\theta^*$).

### 3.1.1   Invariant mass

After the decay/reconstruction variables are determined the algorithm calculates the invariant mass of the two particles:

$$M_{(K\pi)} = \sqrt{(E_- + E_+)^2 - (\vec{P_-} + \vec{P_+})^2} \tag{3.1}$$

with $\vec{P_-}$ and $\vec{P_+}$ the momentum of the positive and negative particle as measured by the detector. The other terms in the equation are assumed to be the energy of the $K^-$ and $\pi^+$:

$$E_- = \sqrt{m_{K^-}^2 + \vec{P_-}^2} \quad \text{and} \quad E_+ = \sqrt{m_{\pi^+}^2 + \vec{P_+}^2} \tag{3.2}$$

where the negative track is assumed to be a kaon and the positive a pion. Only candidate pairs for which the invariant mass is within a certain region around $M_{D^0}$ are accepted. The mass resolution for the $D^0$ reconstruction is about 8 MeV/c². Any combination of tracks for which the invariant mass is outside a window of 24 MeV/c² around the expected $D^0$ mass is rejected.

## 3.2   Cut parameters

After an event is fully reconstructed and the decay parameters determined, two sets of selection criteria are used in order to separate real $D^0$ decay tracks from randomly combined tracks. The first selections are applied at the reconstruction level where tracks are accepted if they were reconstructed with a specific set of detectors (see section 3.6). The second type of selections is implemented at the level of single and two-track property. For single tracks cuts are applied on the transverse momentum ($p_T^K$, $p_T^\pi$) and impact parameters ($d_0^K$, $d_0^\pi$). For a pair of tracks we apply cuts on the distance of closest approach ($dca$), cosine of pointing angle ($\cos\theta_{point}$), cosine of decay angle ($\cos\theta^*$) and product of impact parameters ($d_0^K \times d_0^\pi$).

As described in section 2.4, particles are parametrized as helices which are fully described by 5 parameters: two along the $z$ direction and three in the $xy$ plane. Those alone are enough to determine all other parameters for the two-body decay. Given the finite set of independent variables, all decay parameters will be correlated. We will use those parameters which show little or no correlation and investigate in detail how cuts will affect the selection of the signal.

For very rare processes the signal extraction procedure always faces the difficulty of separating the signal from a large combinatorial background. These processes give only a small signal, comparable with statistical fluctuations in the background. In order to distinguish between a signal peak and a statistical error, one has to make sure the peak is a few times larger than the error. This leads to the definition of the statistical significance, defined as

$$\mathcal{S} = \frac{S}{\sqrt{S+B}} \tag{3.3}$$

with S being the number of entries in the signal and B the number of entries in the background. This gives an objective measure of how well the signal extraction is performed. It tells us how large the signal (S) is compared to the statistical fluctuation of the distribution under the peak (S+B).

The purpose of applying cuts on the reconstructed decay parameters is to optimize the signal and background selections. In order to have a proper estimate, cuts were studied using Monte-Carlo simulations in which the identity of all particles is known. Therefore, the numbers S and B in the previous expression are perfectly determined. Optimal values for the cuts were obtained by maximizing the significance of the signal as a function of the applied cuts.

Two different procedures were used to optimise the significance. The first approach is described in [11] and starts by successively setting cuts, starting with the distance of closest approach between the kaon and the pion. Cuts are optimal if the significance of the $D^0$ signal reaches a maximum. After finding the cut on the *dca*, the cut on the transverse momentum of the kaon and pion was optimized. The newly found cuts were kept together with the *dca* and the cut on the cosine of the decay angle was optimized. The same procedure was performed successively on the $K$ and $\pi$ impact parameters, product of impact parameters and pointing angle.

This method was used for a first approximation of the cut values and does not take into account the fact that all topological and kinematic variables are correlated. A more thorough approach makes use of a multidimensional analysis of the cuts by searching for the maximum significance of the $D^0$ signal taking into account all possible combinations of cuts. In this way correlations between variables are kept and allows a better $D^0$ selection. Both methods will be explained in more details in the following sections.

### 3.2.1 Transverse momentum

The most important observable measured directly by the detector is the transverse momentum of the particles. According to [11] the lowest momentum that the ALICE detector can reconstruct in Pb-Pb collisions is around 100 MeV/c.

The energy $E$ and 3-momentum **p** of a particle in a reference frame can be transformed to another reference frame moving with velocity $\beta_f$ as:

$$\begin{pmatrix} E' \\ p'_\parallel \end{pmatrix} = \begin{pmatrix} \gamma_f & -\gamma_f\beta_f \\ -\gamma_f\beta_f & \gamma_f \end{pmatrix} \begin{pmatrix} E \\ p_\parallel \end{pmatrix} \tag{3.4}$$

with $\gamma_f = (1 - \beta_f^2)^{-1/2}$ and $p_\parallel$ is the component of $\boldsymbol{p}$ parallel to $\beta_f$. Fig. 3.2 (left panel) shows the decay of the $D^0$ in its rest frame. The kaon and pion propagate back-to-back and have the same total momentum p = 861 MeV/c [35]. The angle between the $D^0$ flight direction and one of the decay products is $\theta^*$. Boosted in the laboratory reference frame, the decay transforms as shown in Fig. 3.2 (right panel) where only the longitudinal components of $\boldsymbol{p}$ change. In this particular case $\beta_f = -|\beta_f|$.

It follows from Eq. 3.4 that the new component of the momentum becomes

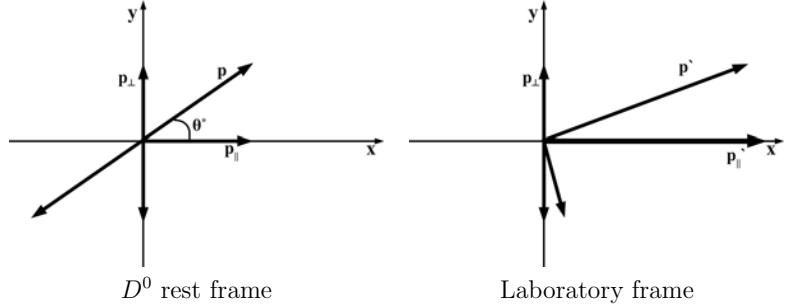

Figure 3.2: $D^0$ rest frame decay (left) boosted to the lab frame (right). Boost transformations are defined in Eq. 3.4

$$p'_\parallel = \gamma_f |\beta_f| E + \gamma_f P \cos \theta^*$$ (3.5)

where E and $P$ are the total energy and momentum of a decay product in the rest frame of the $D^0$. Note that the mass of the particle enters in Eq. 3.5 in the form of energy. For a kaon this results in a slightly larger momentum after the boost. Figure 3.3 shows the transverse momentum distribution of kaons (solid line) and pions (dashed line) originating from a $D^0$ decay.

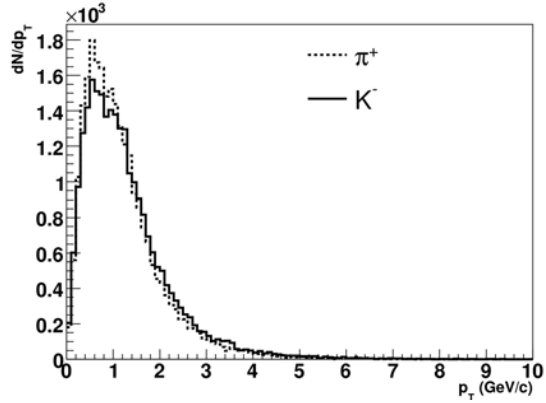

Figure 3.3: $p_T$ spectrum of pions compared to the $p_T$ of kaons - decay products of $D^0$ mesons. Kaons have a slightly larger mean $p_T$ than pions.

### 3.2.2 Decay angle

The cosine of the decay angle is defined as the angle between the kaon and the $D^0$ momentum in the $D^0$ rest frame. Due to the spin 0 of the $D^0$ one expects a uniform emission of the $K^-$ and $\pi^+$ in the rest frame of the meson, corresponding to a uniform distribution of $\cos\theta^*$ ($\frac{dN}{d\cos\theta^*} = 0$). By solving Eq. 3.5 for $\cos\theta^*$ the decay angle becomes:

$$\cos\theta^* = \frac{p'_\parallel - \gamma_f |\beta_f| E}{\gamma_f P} \tag{3.6}$$

where $p'_\parallel$ is the momentum of the kaon measured along the longitudinal flight line of the $D^0$, $E$ is the energy in the rest frame of the $D^0$ and $\beta_f$ is determined by the total $D^0$ momentum in the laboratory frame. Equation 3.5 holds for particles which originate from a $D^0$ decay. For random combinations this equation is not valid.

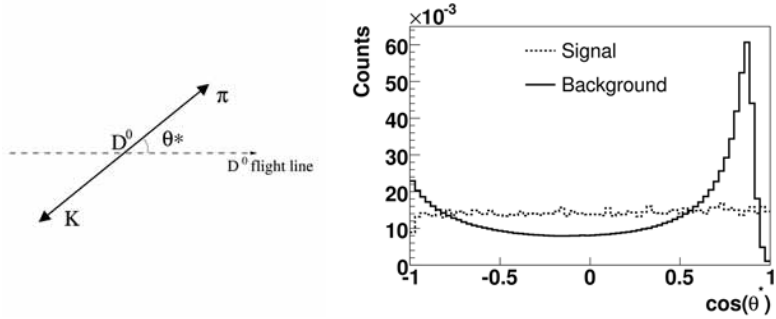

Figure 3.4: Left panel: definition of the decay angle; right panel: $\cos\theta^*$ distribution for background (solid line) and signal (dashed line).

It follows from Fig. 3.4 that the background tends to accumulate at $\cos\theta^* = \pm 1$, thus allowing to set a cut on the decay angle distribution.

### 3.2.3 Impact parameters

The prolongation of a track towards the primary vertex allows for a definition of a signed impact parameter as can be seen in Fig. 3.1. Only the projection on the transverse plane is considered since this gives a more precise measurement than along the $z$ direction.

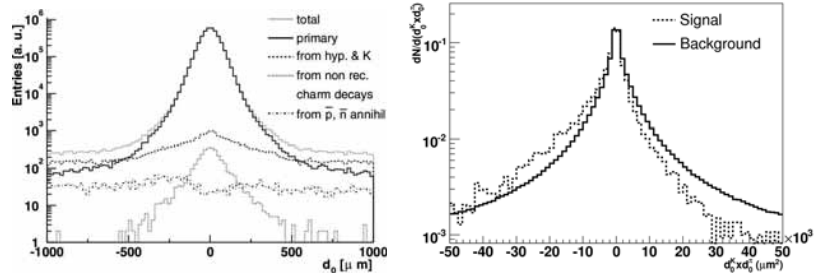

Figure 3.5: Left panel: background sources for impact parameters [11]. Right panel: Product of impact parameters for signal and background pairs with $p_T < 10$ GeV/c - both histograms are normalized to the same integral.

Figure 3.5 (left panel) shows the distribution of different background sources for the impact parameter [11]. For large absolute values the dominant background comes from decays of hyperons and kaons. It would make a legitimate choice for a cut on the impact parameter set as $|d_0| < 500 \mu m$. However, as it will be shown in the next sections, after applying a cut on the transverse momentum, *dca* and cosine of the decay angle the tails of the distributions will be reduced, thus changing the cuts set on the impact parameter as a function of the $D^0$ transverse momentum.

For pairs of tracks originating from a $D^0$ decaying far from the primary vertex the impact parameters should be on opposite sides of the $D^0$ flight line and far from the primary vertex as shown in Fig. 3.1. As a result, the product of the impact parameters should generally yield a large absolute value and have a negative sign. The background instead, consisting of random tracks, is symmetric around 0. Figure 3.5 (right panel) shows the distribution of the product of impact parameters for signal (dashed line) and background combinations (solid line). The product of impact parameters is strongly correlated to the decay length. In order to determine the decay length one has to determine first the secondary vertex. However, due to the fact that the decay length is extremely small and the resolution on the primary vertex is better than on the secondary vertex it is preferable to work with $d_0^K \times d_0^\pi$ which relies only on the primary vertex.

### 3.2.4 Pointing angle

The pointing angle gives a measure of how well the $D^0$ points back to the primary vertex. For tracks coming from a $D^0$ decay the line between the primary and secondary vertex is very close to the summed momenta of the $K^-$ and $\pi^+$. This corresponds to a peak around 1 for the distribution of the cosine of the pointing angle.

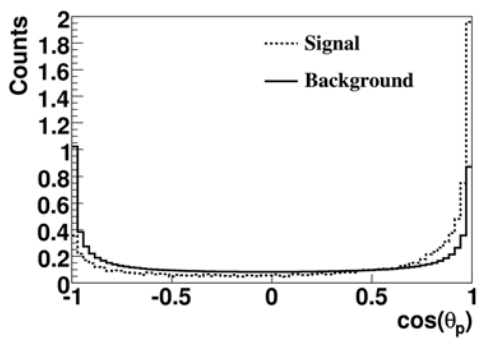

Figure 3.6: Cosine of the pointing angle for signal (dashed line) and background combinations (solid line). Both histograms are normalized to the same integral.

Figure 3.6 shows the distribution of $\cos\theta_p$ for the signal (dashed line). The peak at -1 is due to cases when the secondary vertex is reconstructed on one side of the primary vertex but the total momentum of the kaon and the pion points in the opposite direction. The entries between -1 and 1 are due to intermediate cases, all of which are effects of the secondary vertex resolution. However, a cut placed on the transverse momentum of the kaon and pion reduces the distribution on the negative side, although not completely, while the peak at 1 is less affected. This can be seen in the left panel of Fig. 3.7 (dashed line) where a cut of 0.5 GeV/c is applied on both kaon and pion transverse momentum.

Simulations show that for random combinations the distribution of the cosine of the pointing angle is much more symmetric than the signal and it is peaked at both -1 and 1. Figure 3.6 shows the distribution of $\cos\theta_p$ for background pairs (solid line).

The most important difference between the signal and background distributions is that, if a cut is placed on the kaon and pion transverse momentum, the background is significantly reduced, while the signal is less affected by the cut. Figure 3.7 (upper panel) shows the signal and background distribu-tion with a cut of 0.5 GeV/c on the $K$ and $\pi$ transverse momentum. In the lower panel it is shown the distribution of $\cos\theta_p$ for signal and background between 0.7 and 1, with a cut on $K$ and $\pi$ of 0.8 GeV/c. Notice that the background is significantly reduced compared to the signal.

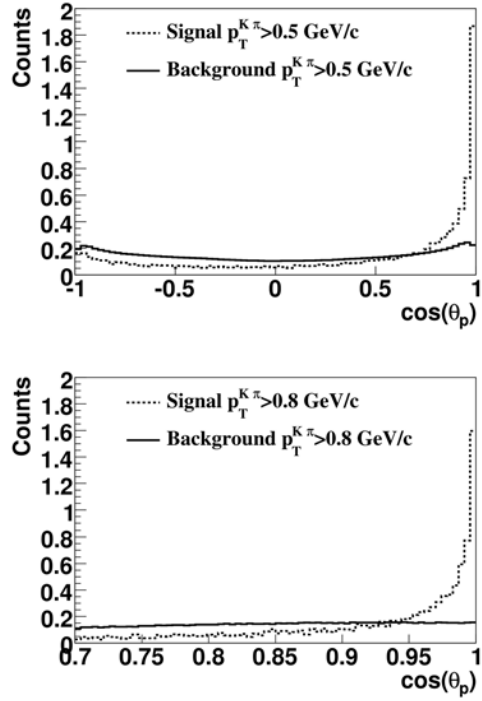

Figure 3.7: $\cos\theta_p$ for signal (dashed line) and background (solid line). Upper panel: a cut on $p_T^{K\pi} > 0.5$ GeV/c is applied. Lower panel: $p_T^{K\pi} > 0.8$ GeV/c.

## 3.3 Cut optimization

In the first part of this section we will focus on the cuts obtained with the linear method also used in [11]. It is applied for 4 bins of $D^0$ transverse momentum: $1 < p_T < 2$, $2 < p_T < 3$, $3 < p_T < 5$, $5 < p_T < 10$ GeV/c. Results will be shown for two $p_T$ intervals in order to show how the cuts behave in different momentum regimes. The obtained value for the optimal cut will be marked with an arrow on the axis of the distribution and the values summarized in Table 3.1. The second part of this section will describe the multidimensional cut analysis method with the cut values summarized in Table 3.2.

### 3.3.1 Linear method

The significance plots shown in the following are functions obtained from the signal and background distribution for various variables. The significance calculation can be expressed as:

$$\mathcal{S}(x) = \frac{\int_{a(x)}^{b(x)} S(t)d\mathbf{t}}{\sqrt{\int_{a(x)}^{b(x)} S(t)d\mathbf{t} + \int_{a(x)}^{b(x)} B(t)d\mathbf{t}}} \tag{3.7}$$

where $x$ is a cut value and the limits of the integrals are expressed as $a(x) = \mathbf{a} + \mathbf{c_1} \cdot x$ and $b(x) = \mathbf{b} - \mathbf{c_2} \cdot x$. $\mathbf{a}$ and $\mathbf{b}$ are the limits of the distributions for the given variable and the coefficients $\mathbf{c_1}$ and $\mathbf{c_2}$ are either 0 or 1 depending on which limit of the integral is fixed (variable). For a cut variable that has to be larger (smaller) than a certain value $\mathbf{c_1}$=1, $\mathbf{c_2}$=0 ($\mathbf{c_1}$=0, $\mathbf{c_2}$=1) and for variables for which the absolute value is calculated $\mathbf{c_1}$=1, $\mathbf{c_2}$=1. For each described cut variable the limits $a(x)$ and $b(x)$ will be explicitly given.

The linear analysis procedure starts by maximizing the $D^0$ significance as a function of the distance of closest approach between the kaon and pion. Pairs for which the *dca* is larger than a certain value are rejected. Fig. 3.8 shows $\mathcal{S}(x)$ for $2 < p_T^{D^0} < 3$ GeV/c and $p_T^{D^0} > 5$ GeV/c. The integration limits for $S$ and $B$ are from 0 to $x$ with $x \in (0, 0.1)$ cm.

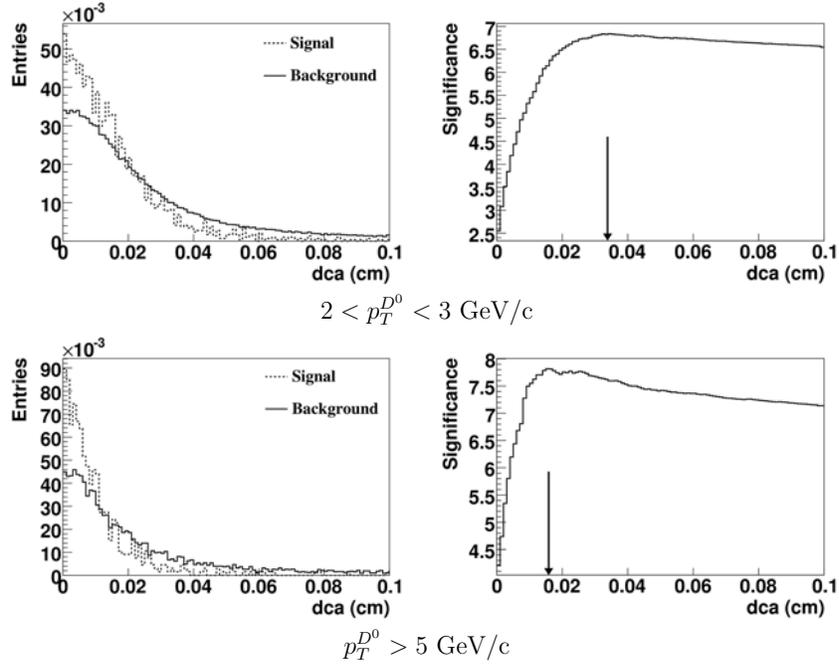

$2 < p_T^{D^0} < 3$ GeV/c

$p_T^{D^0} > 5$ GeV/c

Figure 3.8: Left: distribution of the distance of closest approach. Both histograms are normalized to the same integral. Right: significance of the $D^0$ as a function of the cut on the *dca*.

For $D^0$ mesons with low momentum the combinatorial background is dominated mainly by low energy particles. Therefore it is convenient to place a cut on the two decay daughters in order to filter out the ones with very low momentum. With increasing $D^0$ transverse momentum the average $p_t$ for each of the two decay daughters increases as well. This means that the cut value will also increase. To accommodate the different momenta acquired by the kaon and pion in the decay process, the significance of the $D^0$ meson was calculated separately for $K^-$ and $\pi^+$. Fig. 3.9 (right column) shows $\mathcal{S}(x)$ as a function of the kaon and pion transverse momentum for two $p_T^{D^0}$ intervals. In this case tracks with the transverse momentum larger than a certain value are preferred and so the integration limits are from $x$ to $\mathbf{b}$, with $x$ in the interval 0 to 10 GeV/c. The cut-off at approximately 2 GeV/c (left column) is due to the constraints on the $D^0$ transverse momentum: $1 < p_T^{D^0} < 2$ GeV/c. The cuts on the distance of closest approach have been applied.

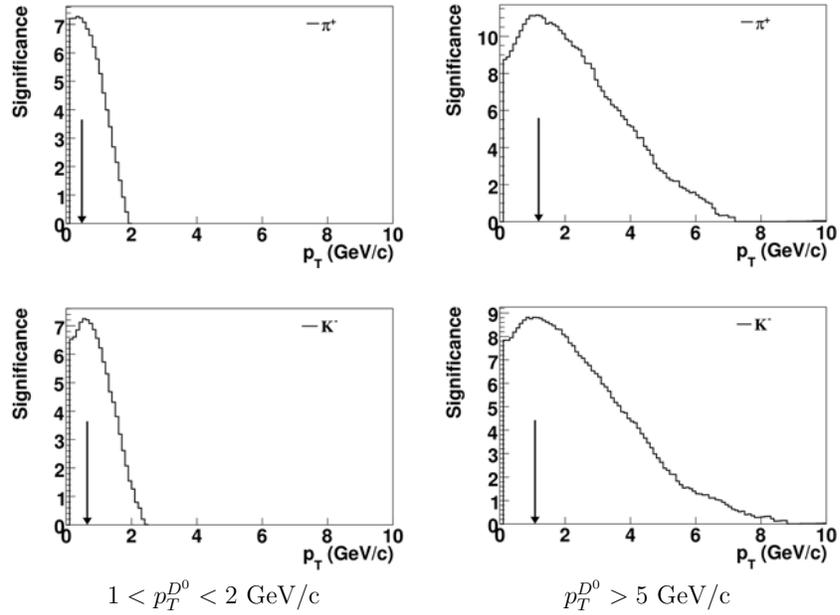

$1 < p_T^{D^0} < 2$ GeV/c $\qquad\qquad p_T^{D^0} > 5$ GeV/c

Figure 3.9: Significance of the $D^0$ as a function of the transverse momentum of the kaon and pion. The cut on the *dca* is applied.

The algorithm proceeds with the significance calculation as a function of the cut on the decay angle. Figure 3.10 shows $\mathcal{S}(x)$ for two $D^0$ transverse momentum bins. In the case of the $\cos\theta^*$ the absolute values are calculated and so the integration limits for $S$ and $B$ are from 0 to 1.

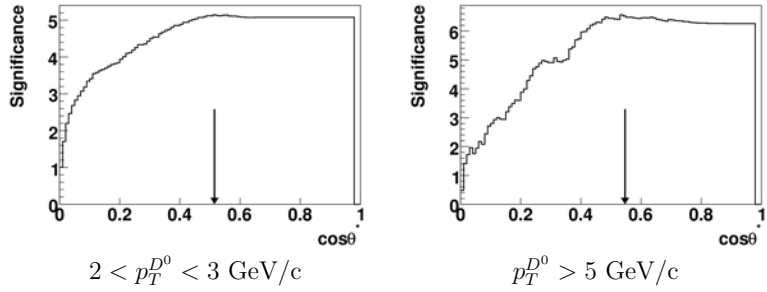

$$2 < p_T^{D^0} < 3 \text{ GeV/c} \qquad\qquad p_T^{D^0} > 5 \text{ GeV/c}$$

Figure 3.10: Significance of the $D^0$ as a function of the cut on the cosine of the decay angle. Previous cuts are applied.

With the increase of the $D^0$ momentum also the impact parameters of the the decay products become smaller. This can be seen if the opening angle between the two tracks in Fig. 3.1 is decreased: the distance to the primary vertex decreases as well. The distribution of the impact parameters for kaons is shown in Fig. 3.11 (left column). The selection on the $D^0$ momentum larger than 5 GeV/c gives a narrower distribution for $d_0^{K^-}$ (bottom-left panel) than at low $p_T$. The impact parameter is required to be smaller than a certain value. Therefore the integration over the signal and background is done in the interval from 0 to $x$ with $x \in [0, 1000]$ $\mu m$.

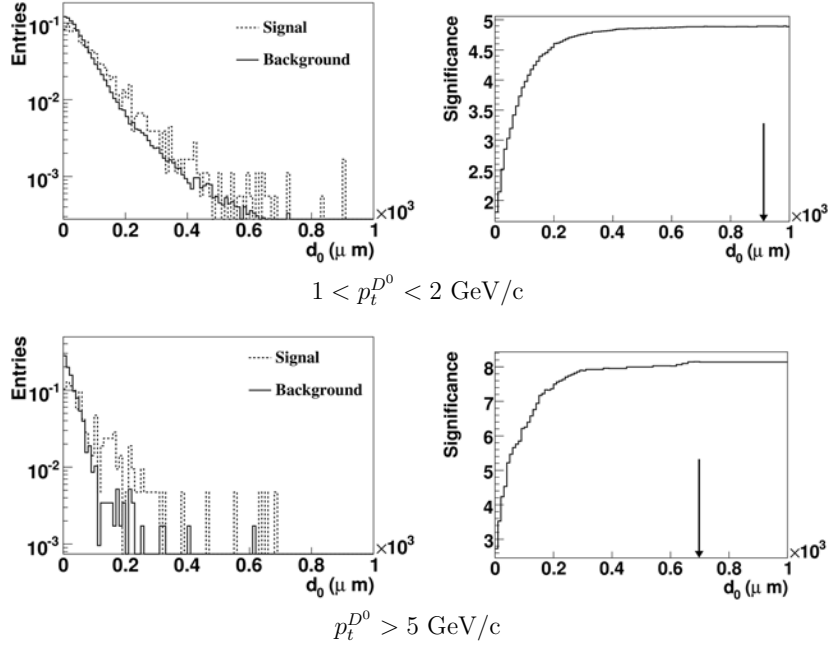

$1 < p_t^{D^0} < 2$ GeV/c

$p_t^{D^0} > 5$ GeV/c

Figure 3.11: Significance of the $D^0$ as a function of the $K^-$ impact parameter. The left column distributions are normalized to the same integral. Previous cuts are applied.

The final cut is the product of impact parameters which is the most powerful cut variable. Figure 3.12 (upper-left panel) shows that at low $D^0$ transverse momentum the signal and background are not well separated. This is because for a low momentum $D^0$ the decay length is very small and therefore the impact parameters for the kaon and the pion are fluctuating around the primary vertex and broadening the distribution for $d_0^K \times d_0^\pi$ around 0. Instead, the higher the momentum of the $D^0$, the more asymmetric the distribution of the product of impact parameters becomes and can be distinguished from the background.

Figure 3.12 shows the behaviour of the product of impact parameters for signal and background for different $D^0$ transverse momentum bins. It can be seen that for $p_T > 2$ GeV/c the cut on the product of impact parameters increases the significance by up to a factor two.

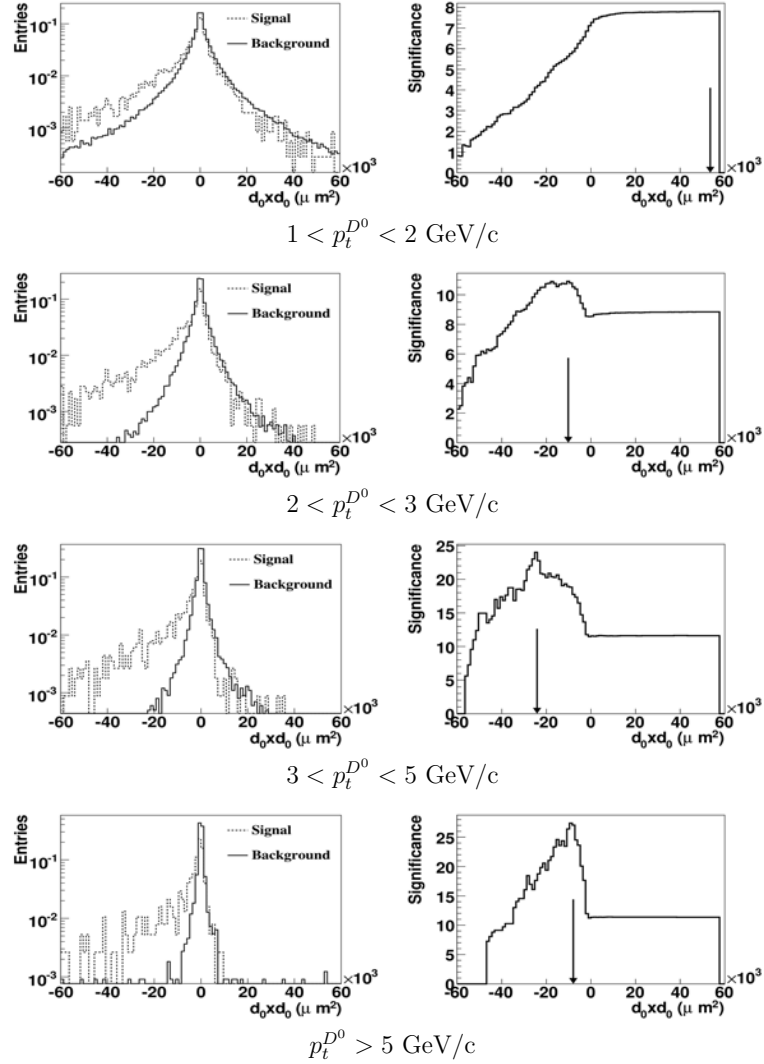

Figure 3.12: Significance of the $D^0$ as a function of the product of impact parameters in various $p_t^{D^0}$ bins. The left column distributions are normalized to the same integral. Previous cuts are applied.

In the last two panels on the right the fluctuation in the significance is due to low statistics in the negative region of the distribution. Nevertheless, the significance shows an increasing trend towards 0 and a sharp drop on the positive side of the distribution.

Figure 3.13 shows the significance of the $D^0$ as a function of the cut on the pointing angle. For tracks coming from a real $D^0$ decay the pointing angle is increasingly peaked at **1** which is reflected by the two plots: the position of the maximum in the significance moves to larger values of $\cos\theta_{point}$ with increasing $p_T$.

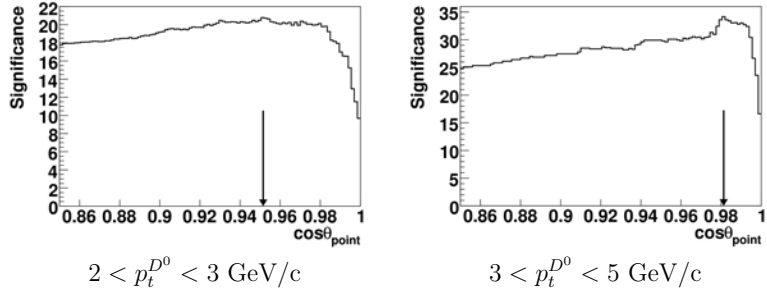

$2 < p_t^{D^0} < 3$ GeV/c $\qquad\qquad$ $3 < p_t^{D^0} < 5$ GeV/c

Figure 3.13: Significance of the $D^0$ as a function of the $\cos\theta_{point}$. Previous cuts are applied.

| **Variable** | $p_T^{D^0} \in [1, 2]$ | $p_T^{D^0} \in (2, 3]$ | $p_T^{D^0} \in (3, 5]$ | $p_T^{D^0} \in (5, 10]$ |
|---|---|---|---|---|
| $dca$ | $< 850\ \mu m$ | $< 350\ \mu m$ | $< 240\ \mu m$ | $< 160\ \mu m$ |
| $p_T^K$ | $> 0.6$ GeV/c | $> 0.7$ GeV/c | $> 0.8$ GeV/c | $> 1.1$ GeV/c |
| $p_T^\pi$ | $> 0.4$ GeV/c | $> 0.6$ GeV/c | $> 1$ GeV/c | $> 1.2$ GeV/c |
| $|cos(\theta^*)|$ | $< 0.68$ | $< 0.52$ | $< 0.58$ | $< 0.55$ |
| $|d_0^K|$ | $< 900\ \mu m$ | $< 650\ \mu m$ | $< 580\ \mu m$ | $< 680\ \mu m$ |
| $|d_0^\pi|$ | $< 900\ \mu m$ | $< 650\ \mu m$ | $< 920\ \mu m$ | $< 770\ \mu m$ |
| $d_0^K \times d_0^\pi$ | $< 56400\ \mu m^2$ | $< -9600\ \mu m^2$ | $< -24000\ \mu m^2$ | $< -9600\ \mu m^2$ |
| $cos(\theta_{point})$ | $> 0.853$ | $> 0.95$ | $> 0.98$ | $> 0.98$ |

Table 3.1: Linear cut analysis in various $p_T$ bins. $D^0$ candidates are kept if the conditions in the table are fulfilled.

### 3.3.2 Multidimensional method

It has been shown in section 3.2.4 that the pointing angle depends on the momentum of both decay daughter tracks and the primary and secondary vertex position. The larger the $D^0$ momentum is the further it decays from the primary vertex and the secondary vertex reconstruction is more accurate. Therefore also the calculation of the pointing angle becomes more accurate. Additionally, the two impact parameters are increasingly localized on opposite sides of the primary vertex. This is shown in Fig. 3.12 where, for increasing $p_T$, the positive tail of the distribution for the product of impact parameters is reduced compared to the negative tail.

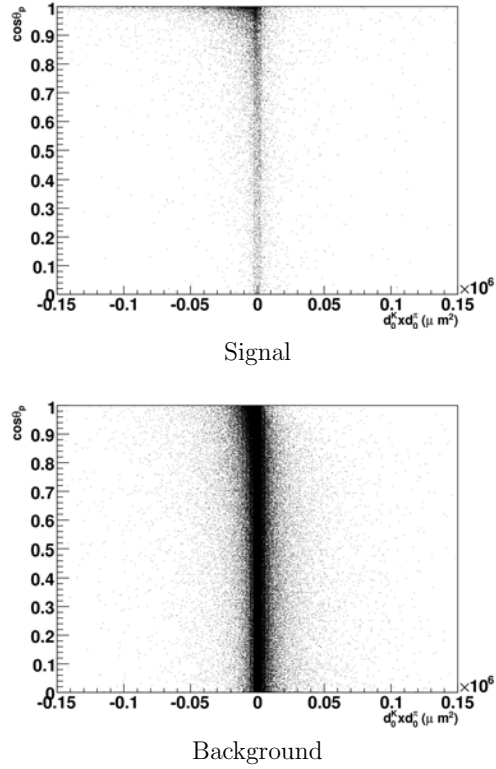

Figure 3.14: Cosine of the pointing angle versus the product of impact parameters for signal and background combinations. A cut on $p_T^{K,\pi} > 0.8$ GeV/c is applied.

Since both product of impact parameters and pointing angle depend on the decay point, it is clear that important correlations exist and the optimum selection for $\cos\theta_{point}$ depends on the cut used for $d_0^K \times d_0^\pi$. If the pointing angle is used in conjunction with the product of impact parameters as a selection criterion it is much stronger than if they are used separately.

Figure 3.14 shows the pointing angle versus the product of impact parameters for signal and background combinations. A cut at 0.8 GeV/c on the transverse momenta of $K$ and $\pi$ was applied. The figure shows that the signal accumulates mainly at negative values for $d_0^K \times d_0^\pi$ and close to 1 for $\cos\theta_p$ while the background is symmetric around 0, with no tail at large $d_0^K \times d_0^\pi$ for $\cos\theta_p$.

In order to incorporate any other correlation between the variables that have been used for the cuts, one has to consider a multidimensional analysis in which the significance of the $D^0$ is calculated using all cuts in parallel. This means an extension of equation (3.7) to an 8-dimensional space and a search for the maximum of the significance:

$$\frac{d}{d\mathbf{x}}\left(\mathcal{S}(\mathbf{x})\right) = 0 \tag{3.8}$$

We will consider the expression $\mathcal{S}$, $\mathbf{S(x)}$ and $\mathbf{B(x)}$ as being matrices with 8 dimensions: $\mathbf{T}^{\mathbf{S}}_{ijklmnop}$ and $\mathbf{T}^{\mathbf{B}}_{ijklmnop}$, treated in C++ as arrays with 8 indices - $p_T^K$, $p_T^\pi$, $dca$, $d_0^K$,$d_0^\pi$, $d_0^K \times d_0^\pi$, $\cos\theta^*$, $\cos\theta_{point}$ - representing the 8 cut variables. The starting and ending cell for each dimension corresponds to the limits of the distribution of each variable. The cells represent the bins of the variables. The filling procedure scans the list of $D^0$ candidates, calculates the cut variables and decides to which bin each of them belongs. In this way the set of variables $\{p_T^K,\ p_T^\pi,\ dca,\ d_0^K,d_0^\pi,\ d_0^K \times d_0^\pi,\ \cos\theta^*,\ \cos\theta_{point}\}$ will be mapped to a set of indices {i,j,k,l,m,n,o,p}. This will determine which cell of the 8-dimensional array will be filled (the content being increased by one). At the end of this procedure each cell contains the number of $D^0$ candidates within a given interval for each of the 8 variables. In order to apply a cut on the $D^0$ it is sufficient to evaluate the sum over the matrix elements from the first indices to a set of indices $\{I, J, K, L, M, N, O, P\}$:

$$S_{IJKLMNOP} = \sum_{i=1}^{I}\sum_{j=1}^{J}\sum_{k=1}^{K}\sum_{l=1}^{L}\sum_{m=1}^{M}\sum_{n=1}^{N}\sum_{o=1}^{O}\sum_{p=1}^{P} T_{ijklmnop} \tag{3.9}$$

By summing up all contents of the matrix the total number of $D^0$ candidates is obtained. If one wants to recover, for example, the distribution of the product of impact parameters one has to sum up the contents of $\mathbf{T}$ for 7 indices. Each cell of the $8^{th}$ index will contain the total amount of $D^0$ candidates with $d_0^K \times d_0^\pi$ within the limits of the given bin.

This filling procedure is applied for both signal and background candidates in order to calculate the significance $\mathcal{S}(\mathbf{x})$.

This algorithm is more powerful than the linear cut method because it keeps all correlations but it has two major deficits: it requires a large amount of computer memory to be allocated for each matrix and it involves also a large number of calculations. The amount of memory required depends on the number of bins allocated for each index of the constructed arrays. The number of cells in a matrix is $\mathbf{N}_b^\mathbf{D}$ - where $\mathbf{N}_b$ is the number of bins in each dimension and $\mathbf{D}$ the number of dimensions. Since $\mathbf{D}$ represents the number of parameters involved for the $D^0$ reconstruction the only "degree of freedom" that is left is the number of bins that can be allocated for each dimension. The maximum number that could be reached was 10 bins for each dimension. The array contains then $10^8$ cells. However, given the previous study of the cuts, we have a rough estimate of the intervals where the significance exhibits a peak. Therefore, by carefully selecting the sensitive regions one can still be accurate enough even with the limitation of the 10 bins per dimension.

After the filling procedure is completed the algorithm proceeds with the calculation of the significance for the $D^0$ as a function of each applied cut.

A straightforward method for the significance calculation would be to calculate all sums expressed in equation 3.9 where all indices { I, J, K, L, M, N, O, P } run from 1 to $\mathbf{N}_b$. However, this implies to repeatedly evaluate a series of sums which in total gives a number of operations in the order of $10^{16}$. In order to avoid calculating the same sums over again, the algorithm runs over each cell of the $T^S$ and $T^B$ matrix and stores the sum over the previous cells. In this way, every time one of the indices is increased, $S_{IJKLMNOP+1}$ does not have to be calculated again but just to be summed with the neighbouring cells. Therefore the number of summations that needs to be done is dramatically reduced, thus significantly reducing the computing time.

The cut values obtained with this procedure are summarized in Table 3.2 where each column gives the value of the cuts for a given $p_T^{D^0}$ interval.

| **Variable** | $p_T^{D^0} \in [1,2]$ | $p_T^{D^0} \in (2,3]$ | $p_T^{D^0} \in (3,5]$ | $p_T^{D^0} \in (5,10]$ |
|---|---|---|---|---|
| $dca$ | $< 280 \ \mu$m | $< 320 \ \mu$m | $< 220 \ \mu$m | $< 200 \ \mu$m |
| $p_T^K$ | $> 0.65$ GeV/c | $> 0.80$ GeV/c | $> 0.60$ GeV/c | $> 1.1$ GeV/c |
| $p_T^{\pi}$ | $> 0.65$ GeV/c | $> 0.80$ GeV/c | $> 0.60$ GeV/c | $> 1.1$ GeV/c |
| $|cos(\theta^*)|$ | $< 0.88$ | $< 0.56$ | $< 0.96$ | $< 0.72$ |
| $|d_0^K|$ | $< 720 \ \mu$m | $< 680 \ \mu$m | $< 800 \ \mu$m | $< 560 \ \mu$m |
| $|d_0^{\pi}|$ | $< 720 \ \mu$m | $< 680 \ \mu$m | $< 800 \ \mu$m | $< 560 \ \mu$m |
| $d_0^K \times d_0^{\pi}$ | $< -40000 \ \mu$m$^2$ | $< -14000 \ \mu$m$^2$ | $< -10000 \ \mu$m$^2$ | $< -6000 \ \mu$m$^2$ |
| $cos(\theta_{point})$ | $> 0.92$ | $> 0.92$ | $> 0.96$ | $> 0.96$ |

Table 3.2: Cut values obtained with the multidimensional analysis for the maximization of the $D^0$ meson significance in various $p_T$ bins. Candidates are kept if the conditions are fulfilled.

## 3.4 $D^0$ significance

The cut values obtained in the previous analysis were applied on the signal and background sample in the four $D^0$ transverse momentum bins. Figure 3.15 shows the $D^0$ significance as a function of the transverse momentum.

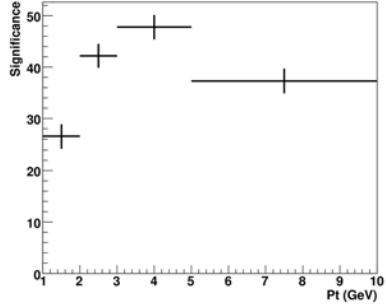

Figure 3.15: $D^0$ significance as a function of $p_T$ with the cuts in Table 3.3 for $10^9$ $pp$ events. Total significance is 78.

It can be seen that the maximum significance ($S_i^{max} = 48$) is obtained in the region $3 \leq p_T^{D^0} \leq 5$ GeV/c which means that this is the most efficient transverse momentum interval for the $D^0$ reconstruction.

| $p_T$ interval | S | B | Significance |
|:---:|:---:|:---:|:---:|
| (1,2) | 1188 | 813 | 26.55 |
| (2,3) | 3312 | 2856 | 42.17 |
| (3,5) | 3276 | 1428 | 47.76 |
| (5,10) | 2268 | 1428 | 37.30 |
| (1,10) | 10044 | 6525 | 78.02 |

Table 3.3: $D^0$ significance for the different $p_T$ intervals normalized to $10^9$ events. The last interval gives the integrated significance.

The significance, as can be seen from Eq. 3.3, is not an additive quantity and has the property that

$$\frac{S_i}{\sqrt{S_i + B_i}} < \frac{\sum S_i}{\sqrt{\sum S_i + \sum B_i}} < \sum \frac{S_i}{\sqrt{S_i + B_i}} \qquad (3.10)$$

## 3.5  $D^{*+}$ reconstruction

The mass of the $D^*(2010)^+$ resonance is only 6 MeV/c$^2$ above the combined mass of the decay products in $D^{*+} \rightarrow D^0 + \pi^+$. The momentum of the two particles in the rest frame of the $D^{*+}$ is:

$$|\mathbf{p}| = \frac{\sqrt{(M^2 - (m_1 + m_2)^2)(M^2 - (m_1 - m_2)^2)}}{2M} = 39 \text{ MeV}/c \qquad (3.11)$$

with $M$ the mass of the $D^{*+}$ and $m_1, m_2$ the mass of the $D^0$ and $\pi^+$, respectively.

In the laboratory frame the process is Lorentz-boosted and the mechanism enhances, on average, the momentum of the heavier particle, as described in section 3.2.1. Figure 3.16 shows that for the pion the Lorentz boost does not increase the $p_T$ by much so that it remains at low momentum and therefore it is referred to as a soft pion.

The momentum of the pion is determined by the decay angle in the rest frame of the $D^{*+}$. The minimum (maximum) momentum for a pion from a $D^{*+}$ decay is when the emission is anti-parallel (parallel) to the flight line:

$$p_{min} = \gamma |\beta| E - \gamma p$$
$$\qquad (3.12)$$
$$p_{max} = \gamma |\beta| E + \gamma p$$

with $\gamma$ and $\beta$ for the $D^{*+}$ in the laboratory frame and $E$ and $p$ the energy/momentum of the pion in the rest frame of the $D^{*+}$. For a $D^{*+}$ with a total momentum of 10 GeV/c, $p_{min} = 522$ MeV/c and $p_{max} = 917$ MeV/c. Figure 3.16 shows the $D^{*+}$ transverse momentum versus the $\pi_{soft}$ transverse momentum.

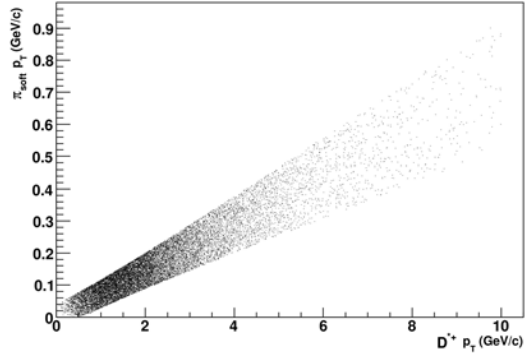

Figure 3.16: Transverse momentum of $D^{*+}$ versus transverse momentum of the soft pion. PYTHIA events with a $D^{*+}$ $p_T$ between 0 and 10 GeV/c.

As described previously, $D^0$ candidates are found by pairing two particles of opposite charge and testing various quality criteria. Whenever such a pair successfully passes as a $D^0$ candidate, the algorithm loops over the new selection of positive particles the invariant mass of the (-++) triplet. This is done independently of the previous $K\pi$ invariant mass calculation and assumes that the third particle is a $\pi^+$:

$$M_{(K\pi\pi)} = \sqrt{(E_- + E_+ + E'_+)^2 - (\vec{P_-} + \vec{P_+} + \vec{P'_+})^2} \qquad (3.13)$$

with $E'_+$ and $\vec{P'_+}$ being the energy and momentum of the third particle.

Because the mass of the $D^0$ is very close to the mass of the $D^{*+}$, it is more advantageous to use the invariant mass difference instead of $M_{D^{*+}}$. This reduces combinatorics for the $D^{*+}$ candidates. $\Delta M = M_{K\pi\pi} - M_{K\pi}$ will measure a value slightly higher than the mass of the $\pi^+$ (a peak at 145 MeV/$c^2$). The uncertainties in the momenta of the kaons and pions cancel out leading to a better resolution for $\Delta M$ than for $M_{D^{*+}}$ [36, 37].

### 3.5.1  $D^{*+}$ significance

The extraction of the $D^{*+}$ is based on the $D^0$ and $\pi_{soft}$ selection. Having the cuts optimized for the maximum $D^0$ significance (section 3.4) one can proceed with the $\Delta M$ spectrum by combining the $D^0$ with the $\pi^+$ candidates.

Integrating the $D^{*+}$ significance on the various $p_T$ bins yields a total significance of 12, calculated in the same way as for the $D^0$. However, the cuts set on the $D^0$ selection are still the ones used for the maximization of the $D^0$ significance. Instead, optimizing the cuts on the $D^0$ for the maximum $D^{*+}$ significance shows that for this procedure the cut values change, especially the product of impact parameters and cosine of pointing angle can be loosened.

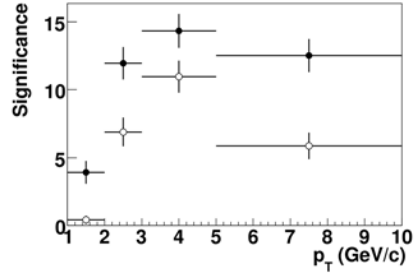

Figure 3.17: $D^{*+}$ significance for the two different cut optimizations. Open symbols for $D^0$ cuts for $D^0$ optimization and solid symbols for $D^0$ cuts optimized for $D^{*+}$ selection. Significance for open symbols is 12 and for solid symbols 22.

| **Variable** | $p_T^{D^0} \in [1,2]$ | $p_T^{D^0} \in (2,3]$ | $p_T^{D^0} \in (3,5]$ | $p_T^{D^0} \in (5,10]$ |
|---|---|---|---|---|
| $dca$ | $< 200\ \mu m$ | $< 400\ \mu m$ | $< 160\ \mu m$ | $< 800\ \mu m$ |
| $p_T^K$ | $> 0.80$ GeV/c | $> 0.80$ GeV/c | $> 1.2$ GeV/c | $> 1.2$ GeV/c |
| $p_T^\pi$ | $> 0.80$ GeV/c | $> 0.80$ GeV/c | $> 1.2$ GeV/c | $> 1.2$ GeV/c |
| $|cos(\theta^*)|$ | $< 0.70$ | $< 0.80$ | $< 0.80$ | $< 1$ |
| $|d_0^K|$ | $< 210\ \mu m$ | $< 350\ \mu m$ | $< 420\ \mu m$ | $< 700\ \mu m$ |
| $|d_0^\pi|$ | $< 210\ \mu m$ | $< 420\ \mu m$ | $< 560\ \mu m$ | $< 700\ \mu m$ |
| $d_0^K \times d_0^\pi$ | $< -20000\ \mu m^2$ | $< -8500\ \mu m^2$ | $< -8500\ \mu m^2$ | $< 10000\ \mu m^2$ |
| $cos(\theta_{point})$ | $> 0.90$ | $> 0.90$ | $> 0.90$ | $> 0.90$ |

Table 3.4: Cut values obtained with the multidimensional analysis for the maximization of the $D^{*+}$ meson significance in various $p_T$ bins. Candidates are kept if the conditions are fulfilled.

Figure 3.17 shows the increase in the $D^{*+}$ significance from one set of cuts, optimized for the $D^0$ selection, to the set of $D^0$ cuts optimized for the $D^{*+}$ selection (open symbols and solid symbols, respectively). The increase in the significance is about a factor 2 and shows that the $D^{*+}$ is less sensitive to the topological selections imposed on the neutral meson. Table 3.4 summarizes the cuts on the $D^0$ optimized for the $D^{*+}$ selection.

The number of reconstructed $D^{*+}$ mesons depends on the efficiency of detecting soft pions. Figure 3.18 shows the $p_T$ distribution of pions from $D^{*+}$ decays compared to the reconstructed $p_T$ of the pions (Monte Carlo information). The majority of the soft pions has a low transverse momentum with the mean around 100 MeV/c while the reconstruction reaches an efficiency of 50% around 250 MeV/c as illustrated in the inset in Fig. 3.18

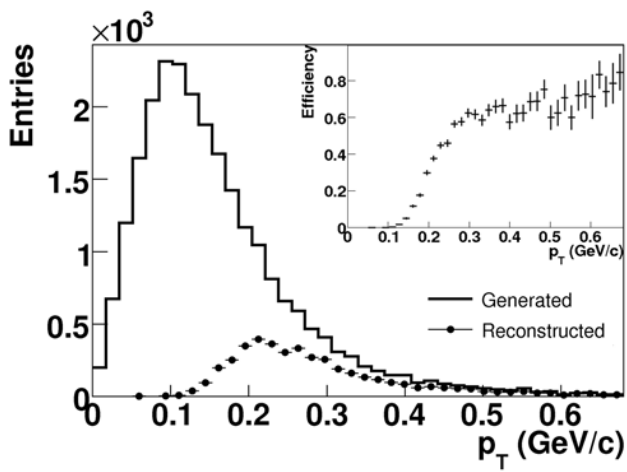

Figure 3.18: Simulated soft pions from $D^{*+}$ decays (solid curve) and reconstructed pions (solid symbols). The inset shows the tracking efficiency.

The tracking is performed with the TPC and ITS sub-detectors combined. As the mean transverse momentum of the $\pi_{soft}$ is lower than the mean transverse momentum in a $pp$ event, the combined TPC and ITS and system is not efficient in this context. In order to increase the $p_T$-range for the reconstructed soft pions we used the Inner Tracking System as a standalone detector as it is better suited for low momentum tracks. This will be discussed in detail in the next section.

## 3.6   ITS-standalone tracking

Previously we discussed the $D^0$ and $D^{*+}$ signal extraction based on the standard reconstruction procedures in ALICE using ITS+TPC combined tracking. Since the experiment is designed to operate with Pb-Pb collisions, detectors were optimized for track recognition in an environment with high track densities. These were predicted to be up to 8000 tracks per unit of rapidity and make track finding a challenging tasks. This will be done both by the Inner Tracking System and the Time Projection Chamber. In the following we will concentrate on the ITS tracking.

As described in ref. [11] the ITS reconstruction software is designed to find prolongations for all tracks found with the TPC. The algorithm is based on the Kalman filter approach with a few adaptations for the two-detector system.

The tracking procedures starts from the outermost pads of the TPC and continues towards the ITS. In this way number of points in the ITS and TPC is assigned to a track. After this step, the remaining hits in the ITS which do not belong to an ITS+TPC reconstructed track are grouped separately and treated by the ITS-standalone procedure. This new set of points in the ITS becomes 3-4 times smaller compared with the initial amount. The ITS-standalone tracking procedure uses the algorithm described in [38] where it is referred to as grouping algorithm.

The general idea of the cluster grouping algorithm is as follows. Since the magnetic field is rather low (0.5 T), tracks are only slightly bent and the $(\theta,\phi)$ coordinates of the clusters belonging to the same track do not change much from one layer of the ITS to another. The angles $\theta$ and $\phi$ are defined in the global coordinate system as

$$\theta = \arctan \frac{z}{\sqrt{x^2+y^2}}$$

$$\phi = \arctan \frac{y}{x}.$$

(3.14)

A schematic view of the hits in the layers of the ITS is shown on the left in Fig. 3.19. A group of reconstructed points found in all layers within the same $(\theta,\phi)$ region is considered a track candidate. This algorithm is especially suited for high-$p_T$ tracks. With small modifications it can be adapted for low momentum tracks as well.

In this case the $(\theta, \phi)$ window is not the same for all layers as shown on the right panel in Fig. 3.19. Given two points in the first two layers of the ITS with the $(\theta,\phi)$ window the track is fitted with a parabola using also the position of the primary vertex. With the first approximation of the track one calculates the position of the next point in the third layer of the ITS. The search is performed in the $(\theta,\phi)$ window around the calculated position. The same procedure is performed for all the layers, using the last three found points for estimating the track curvature. The track candidate is accepted if it has one reconstructed point for each of the layers. In a less restrictive approach, a track candidate is allowed to miss a cluster at one of the layers. The algorithm starts with finding the high-$p_T$ tracks. Then, tracks with decreasing $p_T$ are successively found. The procedure is repeated increasing the size of the $(\theta,\phi)$ window.

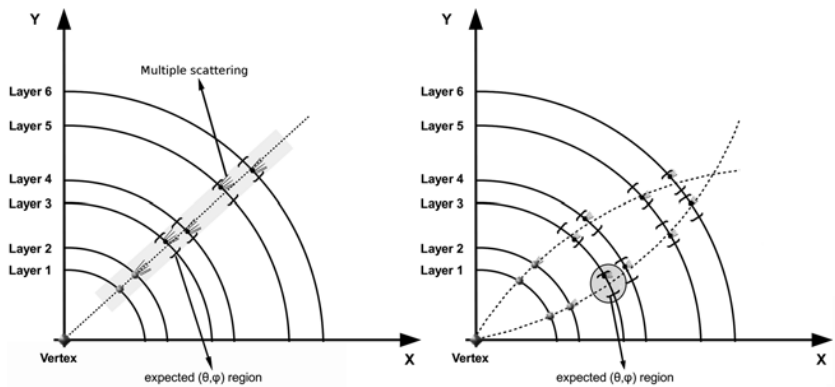

Figure 3.19: Schematic view of tracking in the ITS. Left plot shows the procedure for high-$p_T$ tracks and right for low-$p_T$ tracks.

At the time this thesis was written this algorithm was used in the ITS-standalone reconstruction software. It is regarded as a good tracking procedure although it can be further improved. One of the improvements should consider that the $(\theta,\phi)$ search window should be increased with the layer number. Due to energy loss and multiple scattering in the material, in a given layer, the particle trajectory will be deflected more than in the previous one so the search region should increase in order to find the next hit.

### 3.6.1 Soft pion efficiency

The search windows described previously were initially set to very low values ($\theta_1 \in (0.002, 0.0145), \phi_1 \in (0.003, 0.008)$). In order to improve the reconstruction for lower transverse momentum, the standalone tracking procedure has been tested using several $(\theta,\phi)$ search-windows. This was done by gradually extending the limits of $\theta$ and $\phi$ search intervals up to a factor 4.5 and for $\phi$ by a factor 5.4. The reconstruction efficiency, defined as the number of reconstructed tracks over the number of simulated ($N_{rec}/N_{sim}$), significantly improved from $\sim 16\%$ to 47%.

| $\Theta_i$ | | $\Phi_i$ | | Efficiency | Fraction of fake tracks |
|---|---|---|---|---|---|
| Min | Max | Min | Max | | |
| 0.003 | 0.0080 | 0.002 | 0.01450 | 16.4 % | 1.6% |
| 0.003 | 0.0136 | 0.002 | 0.02726 | 33.2 % | 3.5% |
| 0.003 | 0.0192 | 0.002 | 0.04002 | 40.1 % | 5.1% |
| 0.003 | 0.0248 | 0.002 | 0.05278 | 43.6 % | 6.5% |
| 0.003 | 0.0304 | 0.002 | 0.06554 | 43.7 % | 7.0% |
| 0.003 | 0.0360 | 0.002 | 0.07830 | 47.0 % | 8.8% |
| 0.003 | 0.0640 | 0.002 | 0.14210 | 47.8 % | 13.1% |

Table 3.5: $\theta, \phi$ windows chosen for the tuning of the ITS-standalone reconstruction. The reconstruction efficiency saturates around 47%.

Table 3.5 shows that for the extension of the search intervals by a factor $\sim 1.7$ the reconstruction efficiency already increases by a factor two. The efficiency saturates around 47% for $\theta \in (0.003, 0.036)$, $\phi \in (0.002, 0.0783)$ for large search windows.

The extension of the search intervals, besides improving the soft pion reconstruction efficiency it also introduces so-called 'fake tracks'. A reconstructed track for which more than 20% of the hits that were associated with it belong to another (simulated) track is regarded as fake. Widening the search regions on the ITS layers increases the probability of hits being wrongly assigned to other tracks. When particles pass through the ITS material secondary particles can be created. The secondaries will also leave hits as they further cross the ITS layers. Having a small momentum they are much more curved than real soft pions and hence the hits left in the material will be much more displaced and wrongly associated to other tracks. Simulations have shown that the majority of the fake tracks have decay products which originate from the layers of the ITS. Most of them are delta rays, electrons kicked out of the material. Figure 3.20 shows the position in the $xy$ plane of the starting point of the 'decay products' of fake tracks. It can be seen that the majority has the starting point on the layers of the ITS which suggests that fake tracks have associated hits produced by secondary particles from the ITS material. Points in the interstitial space correspond to decays of the soft pions. Table 3.5 shows also the fraction of the soft pions which are fake.

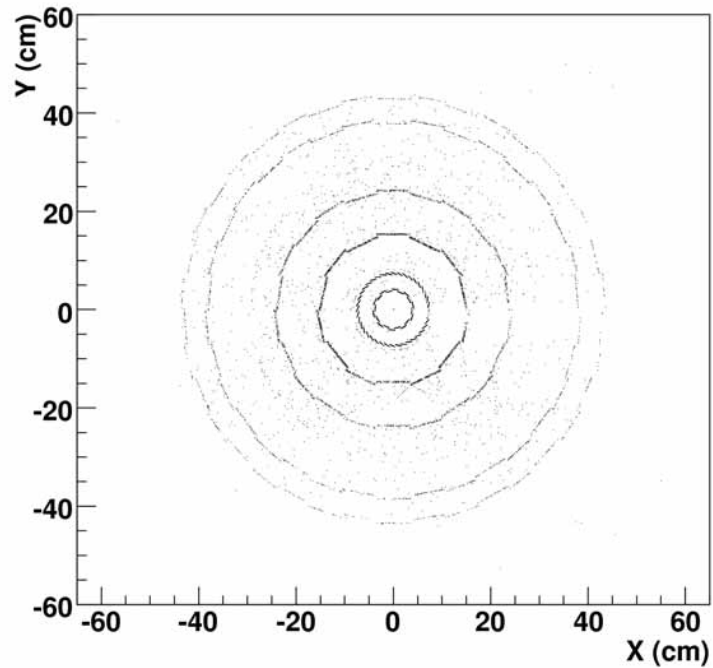

Figure 3.20: XY position in space for the starting point of the 'decay products' of fake tracks corresponding to 25000 soft pions.

Figure 3.21 shows the transverse momentum distribution for generated $\pi^+$ from $D^{*+}$ decays compared with the reconstructed $p_T$ for three $\theta, \phi$ intervals. Solid symbols correspond to the default setup used for Pb-Pb collisions, while the open symbols correspond to $\theta \in (0.003, 0.036), \phi \in (0.002, 0.0783)$ (open circles) and $\theta \in (0.003, 0.064), \phi \in (0.002, 0.1421)$ (open triangles). The values corresponding to the open circles have been chosen as a default setup for the ITS-standalone reconstruction in the case of $pp$ collisions.

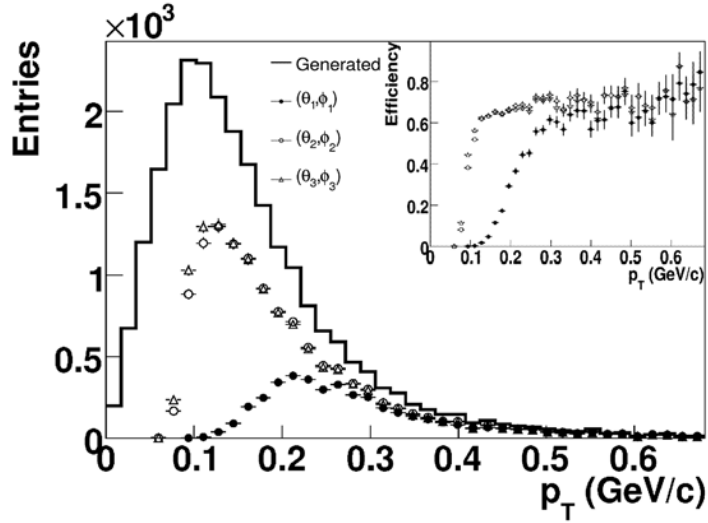

Figure 3.21: Soft pion $p_T$ distribution. Solid line is the generated soft pion spectrum. Solid symbols represent the reconstruction corresponding to the the Pb-Pb optimized ITS tracking - $(\theta_1, \phi_1)$ search window, empty circles correspond to $(\theta_2, \phi_2)$ and empty triangles correspond to $(\theta_3, \phi_3)$.

### 3.6.2   Acceptance efficiency

The production yield for $D^0$ mesons in the pseudorapidity region $|\eta| < 1$ is given in Table 1.3 ($dN/dy_{|y_{lab}|<1} = 0.0196$). In order to understand the total efficiency for the D mesons reconstruction, two classes of efficiencies were introduced: detection efficiency and acceptance efficiency. The detection efficiency was discussed in the previous sections. The acceptance efficiency is defined as the fraction of the produced mesons in the rapidity range $|\eta| < 1$ that have decay products inside the same rapidity region.

For this study one million $D^{*+}$ mesons were generated with PYTHIA in pseudorapidity $|\eta| < 1$ (Fig. 3.22). These decayed (with 68% branching ratio) into $D^0$ and $\pi^+$. The $D^0$ mesons further decayed into $K^- \pi^+$ (3.8% branching ratio). For every decay process in the chain there is a certain probability that the decay product will fall outside the pseudorapidity region of the original particle.

Note that in the decay process of the $D^{*+}$ the $D^0$ meson takes more than 90% of the original momentum (see Fig. 3.16). This means that more $D^0$s will remain in the same $\eta$ as compared to soft pions (see table 3.6). In the case of the $D^0 \rightarrow K^- + \pi^+$ the difference between the masses of the two decay products is not as large and this is also reflected in the relative percentage of kaons and pions remaining in the same pseudorapidity region (see table 3.7).

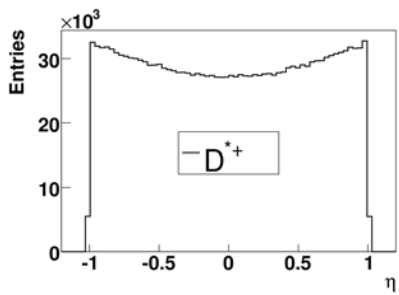

Figure 3.22: Pseudorapidity distribution of 1.04 million $D^{*+}$ mesons. The spectrum is $p_T$-integrated from 0 to 10 GeV/c.

| $D^{*+} \rightarrow D^0 + \pi^+$ (68% BR) | $D^{*+}$ | $D^0$ | $\pi^+_{soft}$ |
|---|---|---|---|
| **Yield** ($|\eta| < 1$) | $10^6$ | 707204 | 645084 |
| **Percentage** ($|\eta| < 1$) | 100% | 99.1% | 90.4% |

Table 3.6: Kinematic efficiencies for the decay $D^{*+} \rightarrow D^0 + \pi^+$. All particles in the decay process have $|\eta| < 1$.

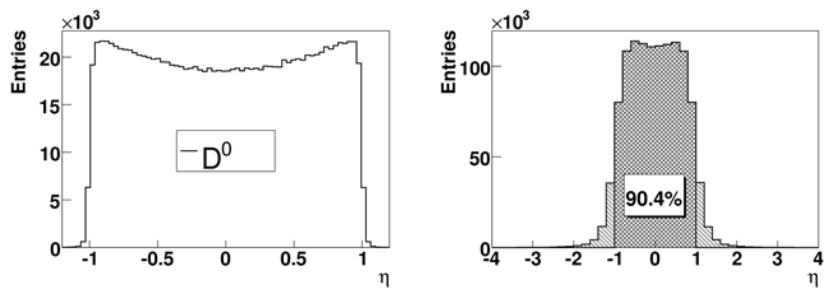

Figure 3.23: $p_T$ integrated $\eta$ distributions of $D^0$ (left) and $\pi^+$ (right) from a $D^{*+} \rightarrow D^0 + \pi^+$ decay.

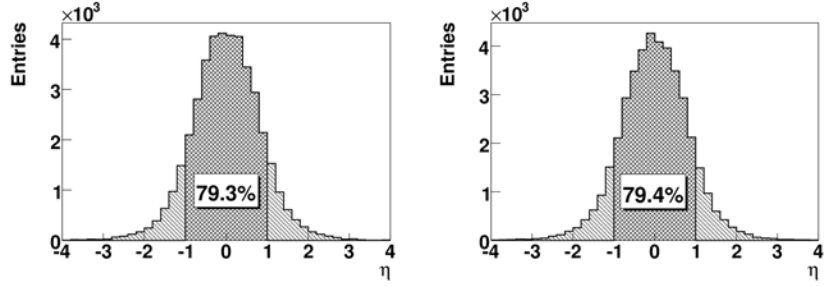

Figure 3.24: $p_T$ integrated pseudorapidity distributions of kaons (left) and pions (right). Decay products from the process $D^0 \to K^- \pi^+$. 79.3% of kaons and 79.4% of pions remain inside $|\eta| < 1$.

| $D^0 \to K^- + \pi^+$ (3.8% BR) | $D^0$ | $K^-$ | $\pi^+$ |
|---|---|---|---|
| **Yield** ($|\eta| < 1$) | $10^6$ | 20386 | 20389 |
| **Percentage** ($|\eta| < 1$) | 100% | 79.3% | 79.4% |

Table 3.7: Acceptance efficiencies for the decay $D^0 \to K^- + \pi^+$. All particles in the decay process have $|\eta| < 1$.

### 3.6.3   $D^0$ and $D^{*+}$ efficiencies

For the study of the detector efficiency all particles which in the previous case were inside the pseudorapidity range $|\eta| < 1$ were run through the reconstruction process. Several quality selections were applied for the reconstructed tracks, thus leading to different efficiencies. In the first set of efficiencies tracks were taken into account only if they were reconstructed by the detector and they were initially in the range $|\eta| < 1$. The obtained efficiencies for single tracks, $D^0$ and $D^{*+}$ mesons are listed in table 3.8. This is the intrinsic reconstruction efficiency of the ALICE detector with the current software implementation.

Figure 3.25 shows the kaon reconstruction efficiencies as a function of the transverse momentum with three types of detector combination: ITS-standalone, TPC+ITS only and TPC+ITS & ITS-SA. An efficiency improvement due to the ITS stand-alone procedure is visible. This is because of the reconstruction of low momentum kaons which decay inside the volume of the ITS and do not leave hits in the TPC. The detection improvement is approximately 11%.

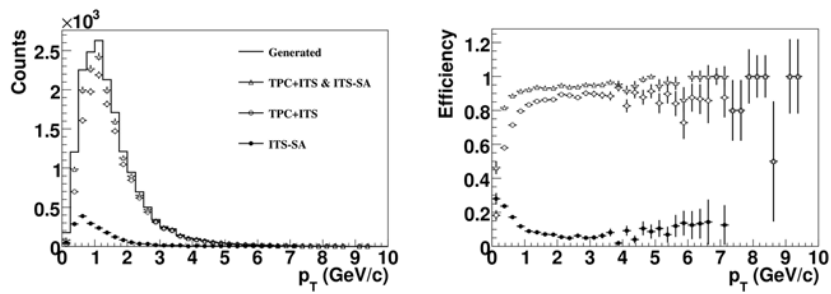

Figure 3.25: Kaon $p_T$ distribution (left) and efficiency (right). Empty triangles: tracks reconstructed with TPC+ITS and ITS-SA. Empty circles: tracks reconstructed with TPC+ITS. Solid circles: tracks reconstructed with ITS-SA only.

In Fig. 3.26 is shown the pion reconstruction efficiency as a function of $p_T$ with the same three detector combinations. The standard TPC+ITS efficiency for pions is in the order of 92%. Pions, having a larger $c\tau$ than kaons, decay less often before reaching the TPC and therefore the ITS-SA procedure increases the reconstruction efficiency by only 3%.

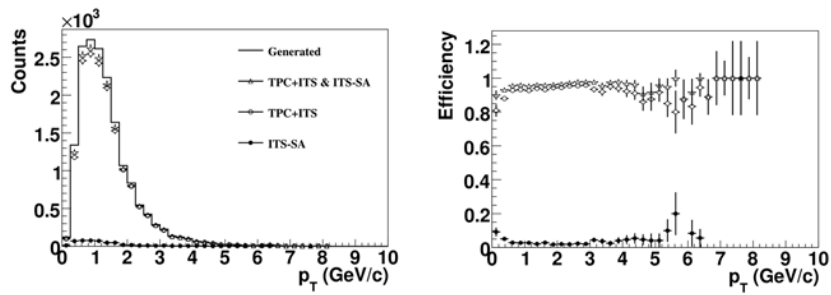

Figure 3.26: Pion $p_T$ distribution (left) and efficiency (right). Empty triangles: TPC+ITS & ITS-SA. Empty circles: TPC+ITS. Solid circles: ITS-SA only.

The combined $K^-$ and $\pi^+$ efficiencies for the $D^0$ reconstruction as a function of the transverse momentum is shown in Fig. 3.27. By accepting kaons and pions reconstructed with TPC+ITS and ITS-SA the $D^0$ efficiency increases by ∼13%.

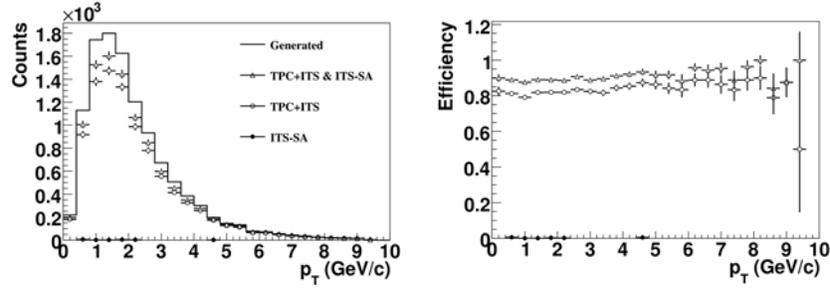

Figure 3.27: $D^0$ transverse momentum distribution (left) and efficiency (right). Open triangles: TPC+ITS & ITS-SA. Open circles: TPC+ITS. Solid circles: ITS-SA only.

The ITS-standalone detection procedure is able to reconstruct 16% of the soft pions. The efficiency as a function of the transverse momentum is shown in Fig. 3.28 for the three types of detector combinations. Due to the increased number of reconstructed soft pions, the detection efficiency for the $D^{*+}$ mesons is also increased, as shown in Fig. 3.29.

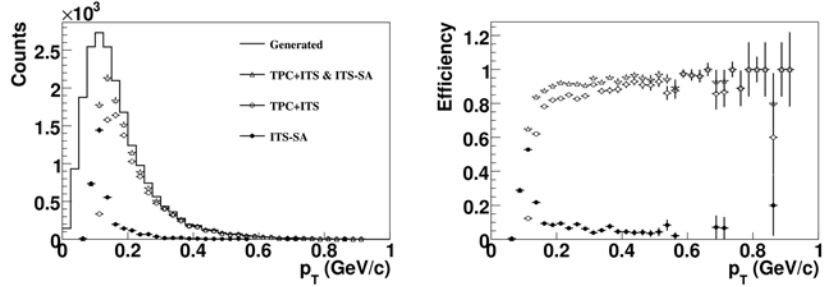

Figure 3.28: Soft pion $p_T$ distribution (left) and efficiency (right). Open triangles: TPC+ITS & ITS-SA. Open circles: TPC+ITS. Solid circles: ITS-SA only.

The overall increase in the $D^0$ and $D^{*+}$ yield is shown in Fig. 3.30. As can be seen, the invariant mass distribution obtained by accepting also tracks reconstructed with the ITS-SA is slightly broader than the one obtained with tracks reconstructed with TPC+ITS only. This is because there are no quality criteria applied on the reconstructed soft pions to illustrate the bare detector efficiency.

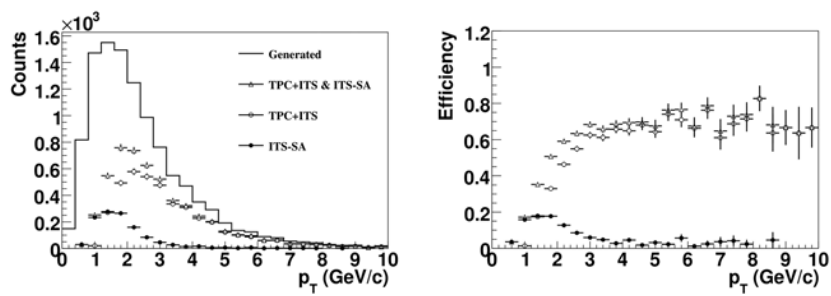

Figure 3.29: $D^{*+}$ transverse momentum distribution (left) and reconstruction efficiency (right). Empty triangles: TPC+ITS & ITS-SA. Empty circles: TPC+ITS. Solid circles: ITS-SA only.

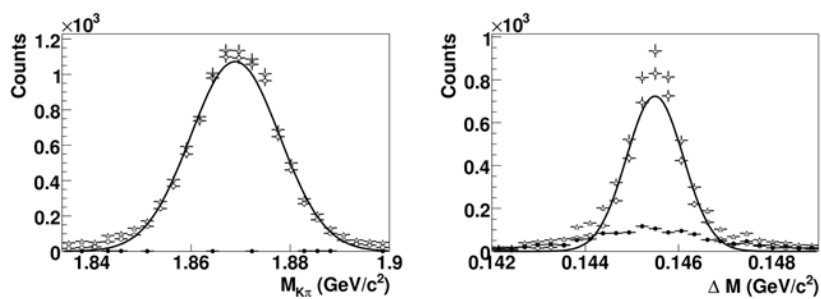

Figure 3.30: Invariant mass distribution for $D^0$ (left) and $M_{D^{*+}} - M_{D^0}$ (right). Empty triangles: TPC+ITS & ITS-SA. Empty circles: TPC+ITS. Solid circles: ITS-SA only. All particles in the decay process have $|\eta| < 1$. No quality selections applied for single tracks.

| Particle type | Reconstruction type | Bare detection efficiency | Quality selection efficiency |
|---|---|---|---|
| $K^-$ | TPC+ITS & ITS-SA | 91.04% | 77.65% |
| | TPC+ITS | 79.67% | 77.65% |
| | ITS-SA only | 11.36% | 0.0% |
| $\pi^+$ | TPC+ITS & ITS-SA | 95.63% | 90.44% |
| | TPC+ITS | 92.69% | 90.44% |
| | ITS-SA only | 2.94% | 0.0 % |
| $D^0$ | TPC+ITS & ITS-SA | 87.04% | 70.12% |
| | TPC+ITS | 73.76% | 70.12% |
| | ITS-SA only | 13.28% | 0.0% |
| $\pi^+_{soft}$ | TPC+ITS & ITS-SA | 65.89% | 65.26% |
| | TPC+ITS | 49.42% | 49.38% |
| | ITS-SA only | 16.46% | 15.88% |
| $D^{*+}$ | TPC+ITS & ITS-SA | 49.0% | 45.90% |
| | TPC+ITS | 37.1% | 34.90% |
| | ITS-SA only | 11.9% | 10.90% |

Table 3.8: Detector efficiencies for single tracks ($K^-$ $\pi^+$ $\pi^+_{soft}$), two-track topology ($D^0$) and three-track topology ($D^{*+}$). All particles have $|\eta| < 1$. $3^{rd}$ column: no quality selections applied for single tracks. $4^{th}$ column: quality selections applied for single tracks - see text below.

To improve the quality of the selected $D^0$ and $D^{*+}$ candidates, several selection criteria were applied on the reconstruction of the single tracks. For the kaon and pion candidates a successful refit was required in the TPC and ITS in order to better determine the momentum of the particles. The refit in the TPC and ITS is done after the track is reconstructed inward from the TPC to ITS (see section 2.4). For the soft pion we require two hits in the pixel detector. The TPC-refit requirement cannot be fulfilled since the particle has a very low transverse momentum and thus does not leave enough hits in the TPC. Another reason why a refit in the TPC cannot be required for soft pions is that a considerable fraction of low momentum pions is reconstructed just with the ITS-standalone mode in which TPC information is missing. Selecting soft pions with a refit in the ITS is also not desirable as this cut is removing almost all pions reconstructed at the lowest momenta. This is shown in Fig. 3.31. Summarized in table 3.8 ($4^{th}$ column) are the efficiencies with these selection criteria applied.

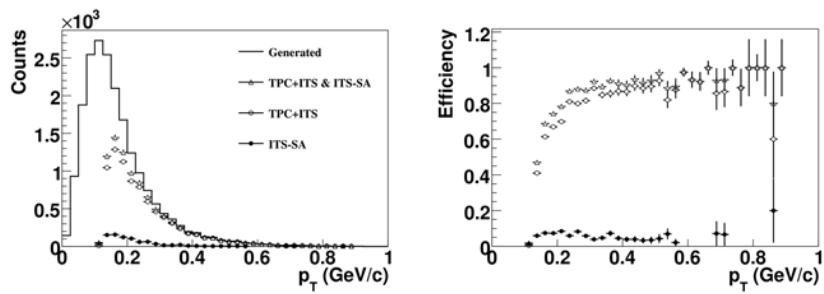

Figure 3.31: Soft pion $p_T$ distribution (left) and efficiency (right). Empty triangles: TPC+ITS & ITS-SA. Empty circles: TPC+ITS. Solid circles: ITS-SA only. A refit in the ITS is required for the soft pion. It removes almost all pions in the low $p_T$ region.

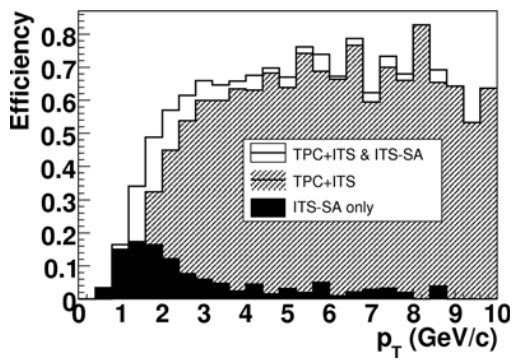

Figure 3.32: $D^{*+}$ efficiency with TPC+ITS & ITS-SA, TPC+ITS and ITS-SA only. Single-track quality cuts were applied: TPCRefit and ITSRefit for $K^-$ and $\pi^+$ and at least two hits in the pixel detector for the $\pi^+_{soft}$.

The increase of the $D^{*+}$ signal from the standard TPC+ITS reconstruction to TPC+ITS & ITS-SA as a function of $p_T^{D^{*+}}$ is shown in Fig. 3.32. The main increase of the efficiency is obtained in the lower region of the transverse momentum where it can go down to 0.6 GeV/c at $\sim 3\%$ efficiency. In the $p_T$ bin from 0.8 to 2 GeV/c the efficiency increases by $\sim 15\%$. Table 3.9 lists the ITS-standalone efficiency in the most significant transverse momentum region.

| Bin edges [GeV/c] | Efficiency |
|:-----------------:|:----------:|
| 0.4-0.8 | 3.4 % |
| 0.8-1.2 | 15.0 % |
| 1.2-1.6 | 17.3 % |
| 1.6-2.0 | 16.4 % |
| 2.0-2.4 | 12.1 % |
| 2.4-2.8 | 7.6 % |
| 2.8-3.2 | 5.9 % |
| 3.2-3.6 | 4.6 % |
| 3.6-4.0 | 2.4 % |

Table 3.9: $D^{*+}$ efficiency with $\pi^+_{soft}$ reconstructed with the ITS-SA.

The decrease in the $D^0$ and $D^{*+}$ yields because of the quality selections applied on the $K^-$, $\pi^+$ and $\pi^+_s$ is illustrated in the invariant mass distribution in Fig. 3.33. The reconstruction efficiencies after these selection are listed in Table 3.8 .

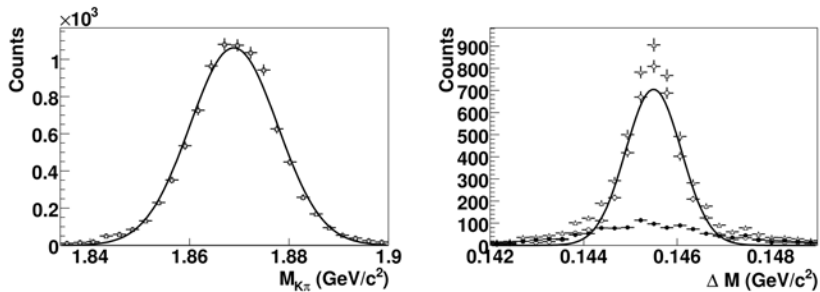

Figure 3.33: Invariant mass distribution for $D^0$ (left) and $M_{D^{*+}} - M_{D^0}$ (right). Empty triangles: TPC+ITS & ITS-SA. Empty circles: TPC+ITS. Solid circles: ITS-SA only. Single-track quality cuts were applied: TPCRefit and ITSRefit for $K^-$ and $\pi^+$ and at least two hits in the pixel detector for the $\pi^+_{soft}$.

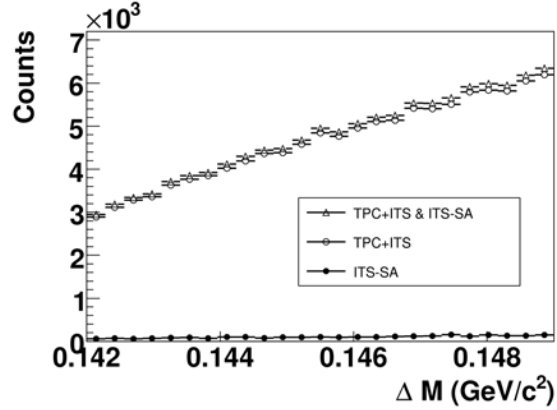

Figure 3.34: Invariant mass difference for the background. Empty triangles: combination of tracks reconstructed with TPC+ITS & ITS-SA. Empty circles: combinations of tracks reconstructed with TPC+ITS. Solid circles: combinations of tracks reconstructed with ITS-SA only. TPCRefit and ITSRefit applied as quality selections for $K^-$ and $\pi^+$ and at least two hits in the pixel detector required for the $\pi^+_{soft}$.

The background under the $D^{*+}$ signal for the two reconstruction procedures (ITS+TPC and ITS+TPS & ITS-SA) is given in Fig. 3.34. For the $p_T$ range between 0 and 10 GeV/c, without any cuts applied, the background increases by a factor 1.02. which is smaller compared to the increase of the signal (factor 1.28). This will lead to an increase in the $D^{*+}$ significance by a factor 1.2.

$$\frac{\mathbf{S}}{\sqrt{\mathbf{S+B}}} \rightarrow \frac{f_S \mathbf{S}}{\sqrt{f_S \mathbf{S} + f_B \mathbf{B}}} \ . \tag{3.15}$$

| Reconstruction | Signif. before cuts | Signif. after cuts |
|---|---|---|
| TPC+ITS & ITS-SA | 15.58 | 31.97 |
| TPC+ITS | 12.96 | 28.66 |

Table 3.10: $D^{*+}$ significance for the different reconstruction types before and after cuts.

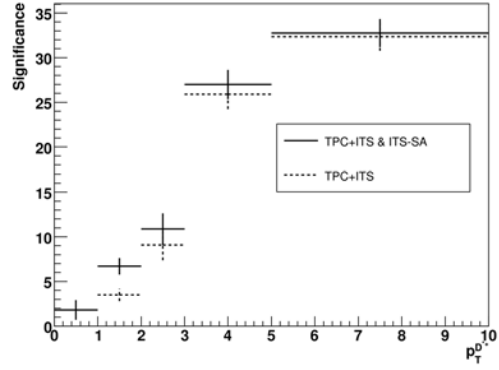

Figure 3.35: Significance as a function of $D^{*+}$ transverse momentum.

| ITS+TPC | | | |
|---|---|---|---|
| $p_T$ interval [GeV/c] | S | B | Significance |
| (0,1) | 0 | 233 | 0 |
| (1,2) | 245 | 4660 | 3.49 |
| (2,3) | 186 | 233 | 9.09 |
| (3,5) | 784 | 132 | 25.89 |
| (5,10) | 2626 | 3961 | 32.35 |
| (0,10) | 3841 | 8986 | 28.66 |
| **ITS+TPC & ITS-SA** | | | |
| (0,1) | 29 | 233 | 1.81 |
| (1,2) | 480 | 4660 | 6.69 |
| (2,3) | 235 | 233 | 10.86 |
| (3,5) | 842 | 132 | 26.98 |
| (5,10) | 2724 | 4194 | 32.75 |
| (0,10) | 4312 | 9452 | 31.97 |

Table 3.11: $D^{*+}$ significance for the different $p_T$ intervals normalized to $10^9$ events.

Figure 3.36 shows the expected significance of the $D^{*+}$ as a function of the number of events. Due to the improved cuts, we expect that with one month of data taking ($\sim 10^8$ events) we are able to detect the $D^{*+}$, up to 10 GeV/c in $p_T$, with a significance of 10. One year of data taking ($10^9$ events) allows us to determine the $D^0$ and $D^{*+}$ transverse momentum distribution up to 10 GeV/c and compare it to theoretical predictions.

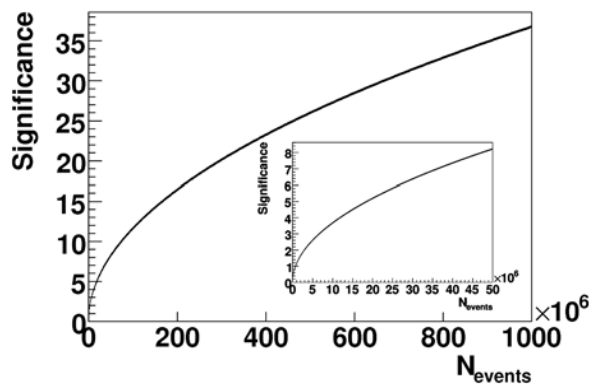

Figure 3.36: $D^{*+}$ significance as a function of number of events.

## 3.7 Sources of systematic errors

In the following section we will give a qualitative discussion about the sources of systematic errors for the $D^0$ and $D^{*+}$ reconstruction and emphasize which are relevant to the present procedure. A quantitative study of the systematic errors extends beyond the scope of this work as it depends on the type of intended physics analysis.

The particle reconstruction consists of a long chain of algorithms and assumptions, each of them being a source of systematic errors and can be categorized as being detector and model dependent.

Sources of systematic errors due to detector performances are:

- Detector acceptance.

- Detector misalignments.

- Tracking performance.

- Primary and secondary vertex resolutions.

- Impact parameter resolution.

- Transverse momentum resolution.

The number of selected $D^0$ and $D^{*+}$ mesons will have to be corrected for efficiency losses introduced by the tracking and selection procedures. Tracking is affected by dead channels and misalignments in the detectors which may lead to a systematic bias in the reconstruction (fake tracks) or to a total loss of the tracks in specific regions. As specified in the previous section, quality selections applied on the single tracks (minimum number of hits in the ITS, a refit in ITS and TPC) should minimize the appearance of fake or badly reconstructed tracks.

The corrections for tracking efficiency and for acceptance are done by means of the Monte Carlo simulation of detector geometry and response. The non-perfect description in the simulation of the geometry and of the physics processes that determine the detector response introduces systematic uncertainties. In ALICE it is considered that these uncertainties will initially be around 10%. However, experience from previous experiments shows that this type of systematic error can be reduced after a few years of running as the understanding of the detector improves.

Algorithms used for particle identification are not a source of concern as PID was not used in the present analysis.

The major sources of systematic uncertainties are the variables which are extensively used for the cuts applied for track kinematics and track pair topology. These include primary and secondary vertex resolutions, as well as transverse momentum resolutions. The most important contributions of these measurements are reflected in the cuts used for the low transverse momentum of the kaon and pion and product of impact parameters for high $D^0$ transverse momenta. The significances obtained for these distributions are the steepest ones and variations in the chosen cut values can influence the number of selected $D^0$ and $D^{*+}$ mesons. Less steep significance distributions should be less affected by the cut value. In this thesis cuts were chosen on the basis of their maximum significance which in some cases show to reduce the signal and background by a large amount. Depending on the type of physics analysis sometimes it is important to have a high $D^0$ purity and sometimes high efficiency. Hence, uncertainties introduced by selection criteria are subject to the analysis purpose and have to be studied accordingly. However, once the detector response is better understood, new Monte-Carlo simulations should improve our knowledge about the stability of the cuts thus reducing the systematic errors introduced in this way.

In addition to the experimental systematic errors, the physics assumptions made in the models for heavy quark production and D mesons yields are also important. The most relevant ones include yields, transverse momentum and rapidity distributions of the $D^0$. Differences between theoretical predictions and measurements taken at a later stage of the experiment have to be accounted and corrected for. This is needed in order to have a more accurate input for the Monte-Carlo generators describing the physics at LHC energies.

# Chapter 4

# Backgrounds

## 4.1 $D^0$ Combinatorial Background

The entire invariant mass distribution is obtained by pairing together all tracks in an event. If the event contains a $D^0$ or a $\overline{D^0}$ then, by coupling a negative particle with a positive one (assuming it is a $K^-\pi^+$ or $K^+\pi^-$ pair), the correct combination will occur. The invariant mass calculation is performed for all $K^\mp\pi^\pm$ combinations, the obtained spectrum consisting of :

1. Uncorrelated background tracks,

2. Correlated real $K\pi$ tracks,

3. Correlated but misidentified tracks.

The sum of the three previous types of pairs forms the distribution shown in Fig. 4.1 as open symbols. The main contribution to the background comes from uncorrelated tracks. The shape of this background can be approximated by using the technique of the "like-sign" background (**LS**) which allows the subtraction of uncorrelated pairs from the "opposite-sign" distribution (**OS**) by using combinations of $K^-\pi^-$ and $K^+\pi^+$ pairs from the same event (solid line in Fig. 4.1)

This method is based on two properties of large amounts of events: the transverse momentum distribution is the same for negative and positive charged particles and the number of tracks is similar for both charges. This is shown in Fig. 4.2 where the transverse momentum distributions for charged particles are almost identical.

However, the inset in Fig. 4.2 shows that on an event-by-event basis the number of positive and negative particles fluctuates. In $pp$ collisions the two numbers can differ by a relatively large amount. The figure shows the distribution of $\Delta N = P - N$ (amount of positive tracks minus amount of negative tracks).

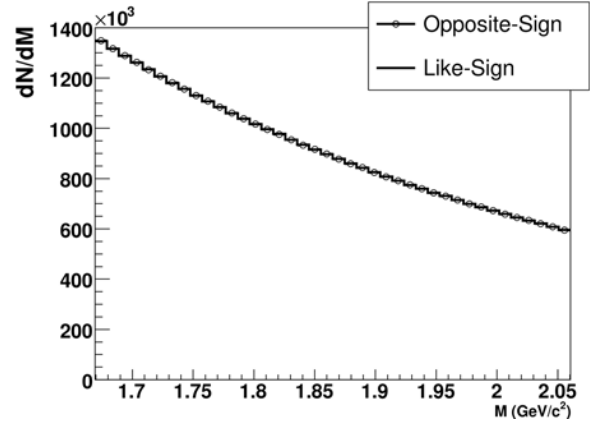

Figure 4.1: Opposite-sign (open symbols) and like-sign (solid line) invariant mass distribution after normalization for $3.3 \times 10^6$ events.

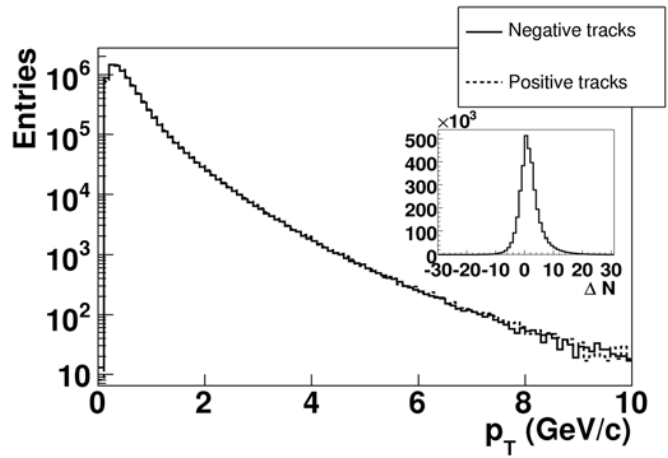

Figure 4.2: Transverse momentum of positive and negative charged particles. The inset shows the distribution of the difference between the number of positive and negative charged tracks on an event-by-event basis.

For large number of events the amount of $K^-\pi^-$ and $K^+\pi^+$ combinations yields almost the same number as the combinations $K^-\pi^+$ and $K^+\pi^-$. The like-sign distribution is calculated using the geometric mean of $N_{K^-\pi^-}$ and $N_{K^+\pi^+}$ as described in [39]:

$$N_{\mathbf{LS}}(m) = 2 \times \sqrt{N_{K^-\pi^-}(m) \times N_{K^+\pi^+}(m)} \qquad (4.1)$$

where N is the number of entries in a bin with the center at the $K\pi$ invariant mass $m$. The like-sign and opposite sign invariant mass distributions are shown in Fig. 4.1. The like-sign spectrum can then be subtracted from the opposite-sign distribution:

$$N_{D^0}(m) = N_{K^-\pi^+}(m) + N_{K^+\pi^-}(m) - N_{\mathbf{LS}}(m). \qquad (4.2)$$

To illustrate the like-sign method, a data sample of 500.000 $pp$ events was analysed. Each event contained one $c\bar{c}$ pair and the decay of the $D^0$ was forced to hadronic decays. Figure 4.3 (left column) shows the invariant mass distribution for opposite-sign (open symbols) and like-sign (solid symbols) combinations in four transverse momentum bins. No cuts are applied on the candidates. The subtraction of the like-sign background is shown in the right pads. Figure 4.4 is similar to Fig. 4.3 but with all selection criteria applied. For $D^0$ transverse momenta between 1 and 2 GeV/c, where most of the $D^0$ candidates are, the selection criteria are less effective than for higher momenta. This is illustrated by the height of the peak relative to the background after the subtraction of the like-sign combinations.

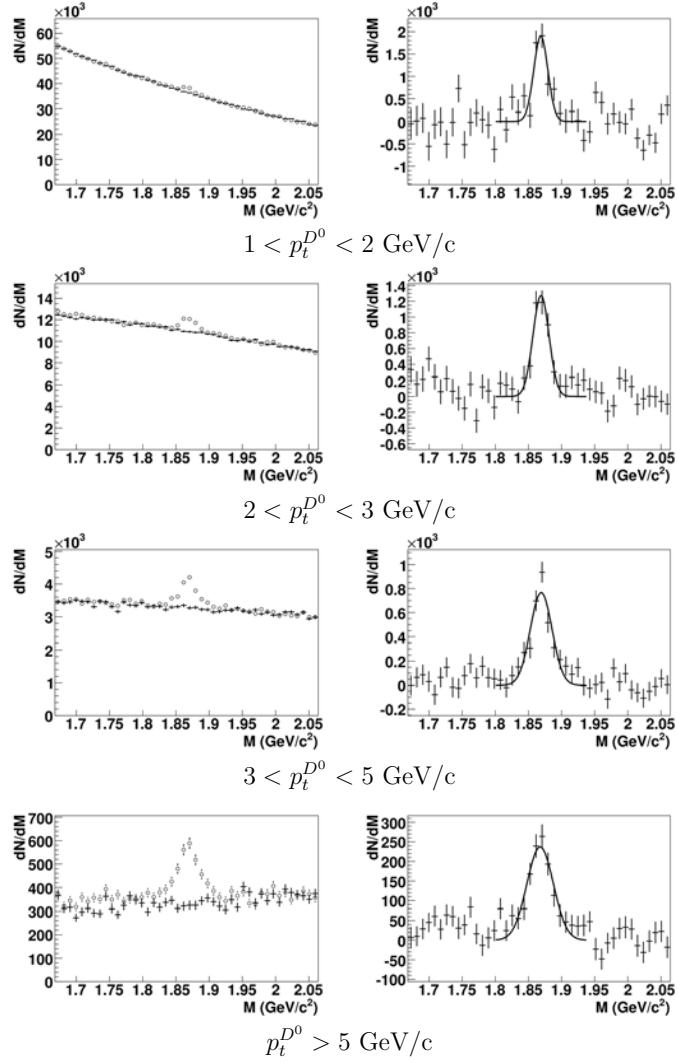

$1 < p_t^{D^0} < 2$ GeV/c

$2 < p_t^{D^0} < 3$ GeV/c

$3 < p_t^{D^0} < 5$ GeV/c

$p_t^{D^0} > 5$ GeV/c

Figure 4.3: Left column: $K\pi$ invariant mass distribution without cuts. Open symbols represent opposite-sign combinations while solid symbols represent like-sign combinations. Right column: $K\pi$ invariant mass distribution after the like-sign background subtraction.

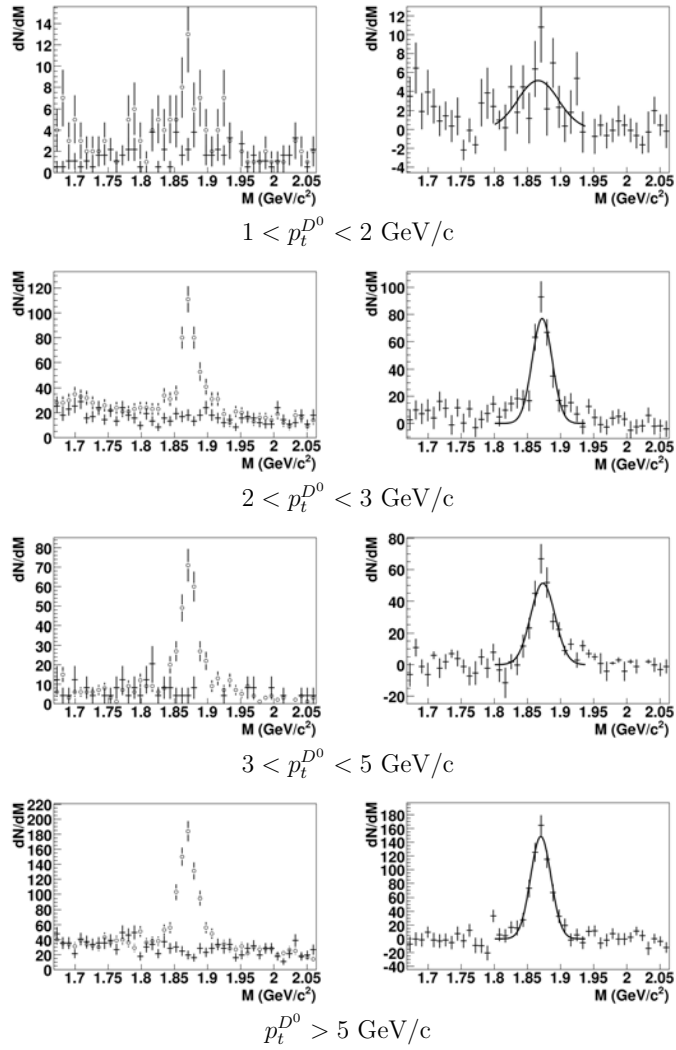

$$1 < p_t^{D^0} < 2 \text{ GeV/c}$$

$$2 < p_t^{D^0} < 3 \text{ GeV/c}$$

$$3 < p_t^{D^0} < 5 \text{ GeV/c}$$

$$p_t^{D^0} > 5 \text{ GeV/c}$$

Figure 4.4: Same as Fig. 4.3 after applying the $D^0$ cuts.

## 4.2 Reflected $D^0$ signal

By pairing all tracks as $K^{\mp}\pi^{\pm}$ there are cases when the kaon is misidentified as a pion and vice versa. This causes a systematic contribution to the invariant mass spectrum of correlated but misidentified particles. The assignment of the kaon mass to the pion leads to a change in the assumed particle energy:

$$E'_- = \sqrt{m_\pi^2 + \vec{P_-}^2} \quad \text{and} \quad E'_+ = \sqrt{m_K^2 + P_+^2} \tag{4.3}$$

Using $E'_-$ and $E'_+$ as the new values, the invariant mass distribution will be centred at the $D^0$ peak but with a much broader peak. This additional correlation in the invariant mass distribution is called the reflected signal. The width of the $D^0$ peak does not depend on the momentum, while the width of the reflection is momentum dependent. Figure 4.5 shows the transverse-momentum dependence of the width of the reconstructed $D^0$ (left panel) compared to the width of the reflected signal (right panel). The contribution of the reflection to the $D^0$ signal varies with the transverse momentum and can range from 20% to 30% as shown in Fig. 4.6.

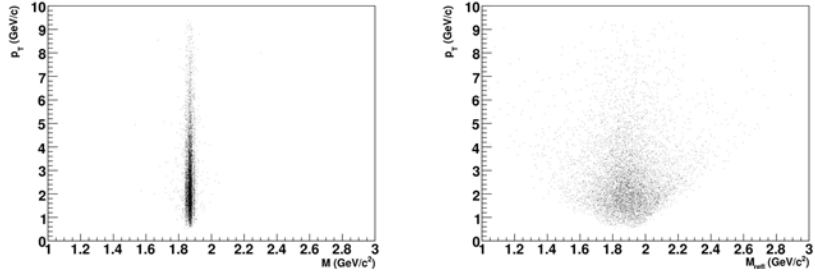

Figure 4.5: $D^0$ invariant mass as a function of the transverse momentum (left panel) compared with the reflection invariant mass as a function of the transverse momentum (right panel).

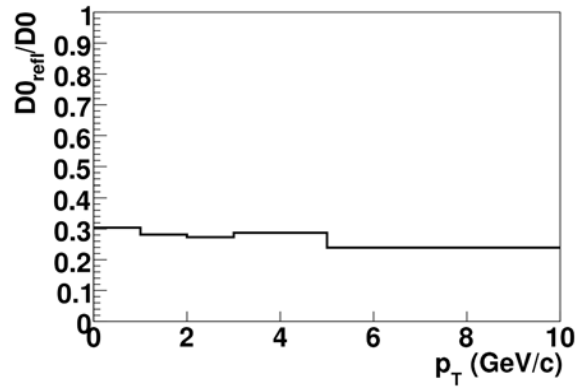

Figure 4.6: The ratio $D^0_{refl}/D^0$ as a function of the $D^0$ transverse momentum after cuts are applied in a $3\sigma$ mass window around the peak.

## 4.3  $D^{*+}$ Background

The shape of the $D^{*+}$ background can be estimated by using the method of $D^0$ side-band combinatorics. This method uses $K\pi\pi$ combinations when the invariant mass of the $K\pi$ pair falls in a region from $6\sigma$ to $10\sigma$ outside the $D^0$ invariant mass peak to estimate the background in the $\Delta M$ distribution for the $D^{*+}$.

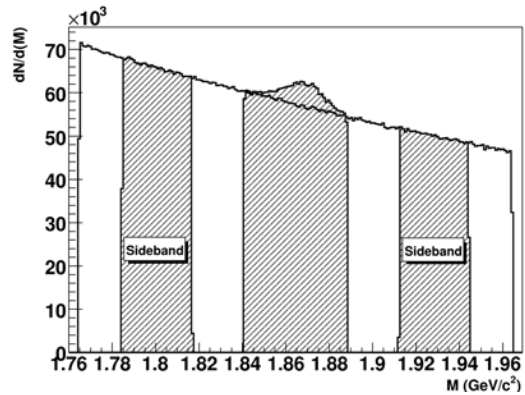

Figure 4.7: Illustration of the $D^0$ side-band selection.

Figure 4.8 shows Monte-Carlo simulations for the background distribution of $\Delta M$ with and without side-band selection. The normalization factor is obtained by integrating both histograms in a region outside the peak and dividing the distribution of $\Delta M$ (open symbols) by the distribution of $\Delta M_{selection}$ (solid line). After the side-band selection and normalization the shape of the selected $\Delta M$ distribution is in excellent agreement with the entire $\Delta M$ distribution.

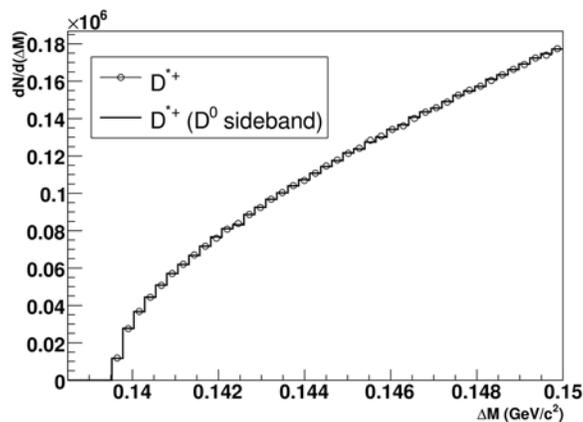

Figure 4.8: $\Delta M$ (open symbols) and $\Delta M$ (solid curve) with $D^0$ side-band selection after normalization.

We applied this method to the $c\bar{c}$ enriched events to illustrate how it performs for a data sample with signal. Figure 4.9 (left column) shows all combinations for $\Delta M$ (open symbols) and $\Delta M$ with the side-band selection on the $D^0$ (points). The right column shows the distribution of $\Delta M$ after the side-band subtraction without applying cuts on the $D^0$ and $D^{*+}$ reconstruction. The distributions in Fig. 4.10 are obtained by applying cuts on the reconstructed $D^0$ and $D^{*+}$ candidates shown in Fig. 4.9. Due to the limited number of events and the tight cuts, the number of entries is significantly reduced. It can be seen in both figures that, in the transverse momentum region between 2 and 3 GeV/c, where the signal is small compared to the background, the side-band subtraction performs a very good signal selection.

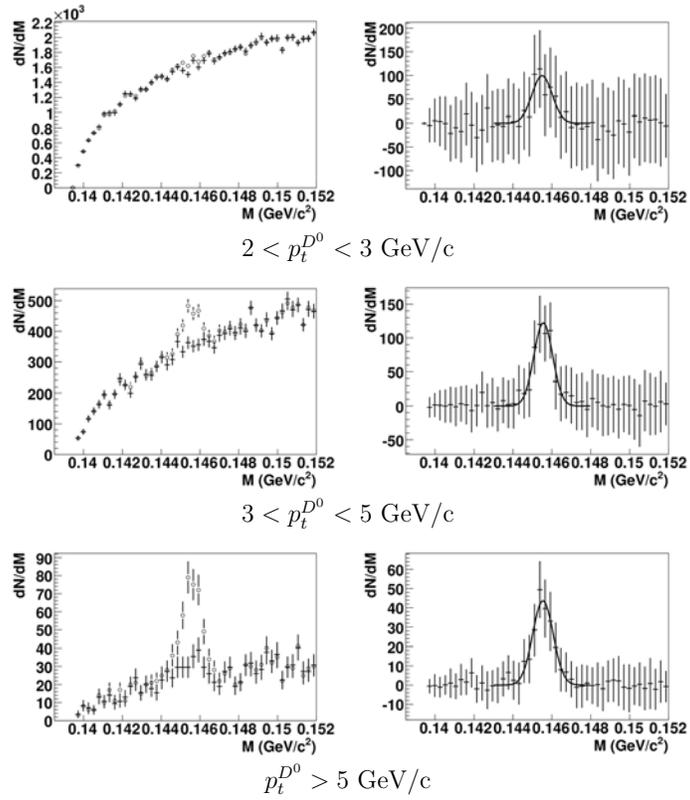

Figure 4.9: Left column: $\Delta M$ distribution (open symbols) and $\Delta M$ with side-band selection (points). Right column: signal after side-band background subtraction. No cuts applied.

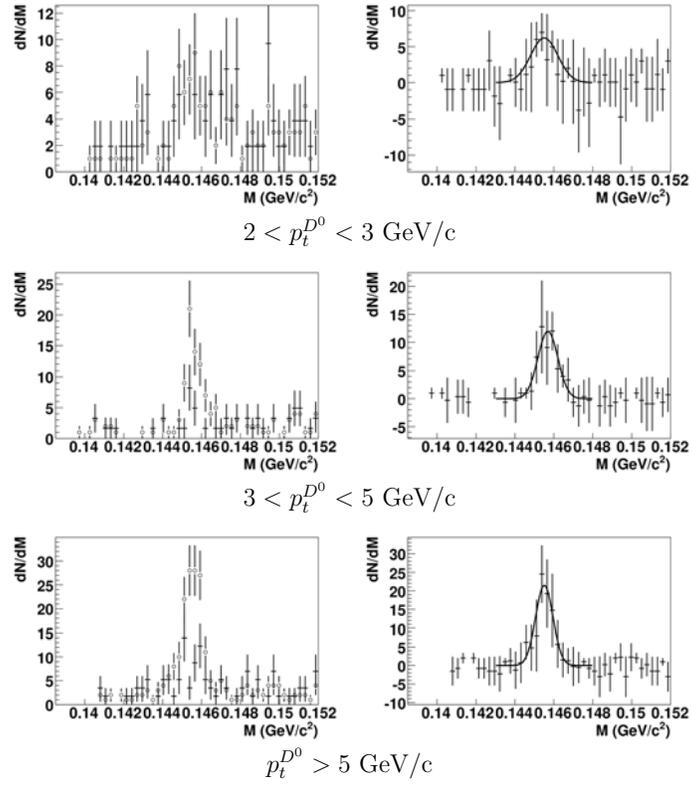

Figure 4.10: Same as Fig. 4.9 after applying the $D^{*+}$ cuts.

# Chapter 5

# Feed-down from beauty decays

## 5.1 Yields and distributions

The goal of the B meson study in the context of this works is to evaluate the fraction of reconstructed $D^0$ mesons which originate from beauty decays. We will show that after applying the $D^0$ selections the initial ratio of reconstructed $(D^0$ from b$)/(D^0$ from c$)$ is enhanced.

The underlying processes for the $D^0$ meson production are summarized as follows:

1. $D^0$ direct production from charm quarks,

2. $D^0$ from $D^*$ decays,

3. $D^0$ from B mesons decays.

Charm fragmentation to $D^0$ and $D^{*+}$ mesons has been discussed in Chapter 1 where the average $D^0$ density at mid-rapidity was reported to be $\langle dN/dy \rangle_{|y_{lab}|<1} = 0.0196$ (including $D^0$ from $D^{*+}$ decays).

Due to its high mass, the b quark is produced in smaller quantities than charm. It is expected that, at the LHC energies, the ratio of the production cross sections for beauty and for charm is around 5% [11]. The fraction of $(D^0$ from b$)/(D^0$ from c$)$ can be calculated as:

$$\frac{dN(b \to B \to D^0)/dy}{dN(c \to D^0)/dy} = \frac{\frac{dN(b \to B)}{dy} \times \mathcal{B}(B \to D^0)}{\frac{dN(c \to D^0)}{dy}} \qquad (5.1)$$

In order to estimate the inclusive branching fractions of B mesons to $D^0$, PYTHIA simulations have been performed considering the following channels: $\overline{B^0} \to D^0$, $B^- \to D^0$ and $\overline{B_s^0} \to D^0$. The intermediate processes, where a B meson decays to an excited D meson, are illustrated in Fig. 5.1. $D^{**}$ are the sums of all higher mass states of the $D^*$.

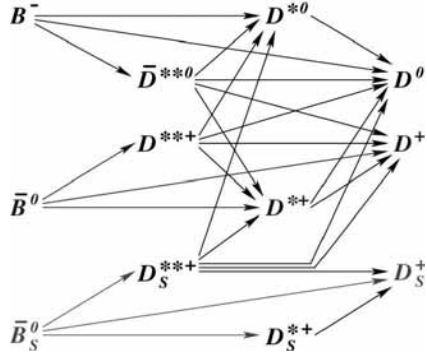

Figure 5.1: B mesons sources of D mesons, taken from [40].

The simulation yields the following inclusive branching fractions:

$$\mathcal{B}(\overline{B^0} \longrightarrow D^0) = 43.40\%$$

$$\mathcal{B}(B^- \longrightarrow D^0) = 89.57\% \tag{5.2}$$

$$\mathcal{B}(\overline{B_s^0} \longrightarrow D^0) = 8.65\%$$

The total yields and mid-rapidity densities for hadrons with beauty are listed in Table 5.1:

| Particle | Yield | $\langle dN/dy \rangle_{|y_{lab}|<1}$ |
|---|---|---|
| $B^0 + \overline{B^0}$ | 0.00577 | 0.00084 |
| $B^+ + B^-$ | 0.00576 | 0.00083 |
| $B_s^0 + \overline{B_s^0}$ | 0.00168 | 0.00025 |
| $\Lambda_b^+ + \Lambda_b^-$ | 0.00106 | 0.00016 |

Table 5.1: Total yield, average rapidity density for $|y| < 1$ and relative abundance for hadrons with beauty in pp collisions at $\sqrt{s} = 14$ TeV [11].

To calculate the $D^0$ yield at mid-rapidity, a set of simulations was dedicated to study the percentage of B mesons producing a $D^0$ outside and inside the ALICE rapidity range ($|y| < 1$). 60.000 $B^-$, $\overline{B_s^0}$ and $\overline{B^0}$ in the rapidity region $|y| < 2$ have been produced. Figure 5.1 shows the decay processes of the B mesons which contribute to the $D^0$ yield. The rapidity and pseudo-rapidity distributions for the generated B and $D^0$ mesons, integrated in the range $0 < p_T^B < 30$ GeV/c, are shown in Fig. 5.2.

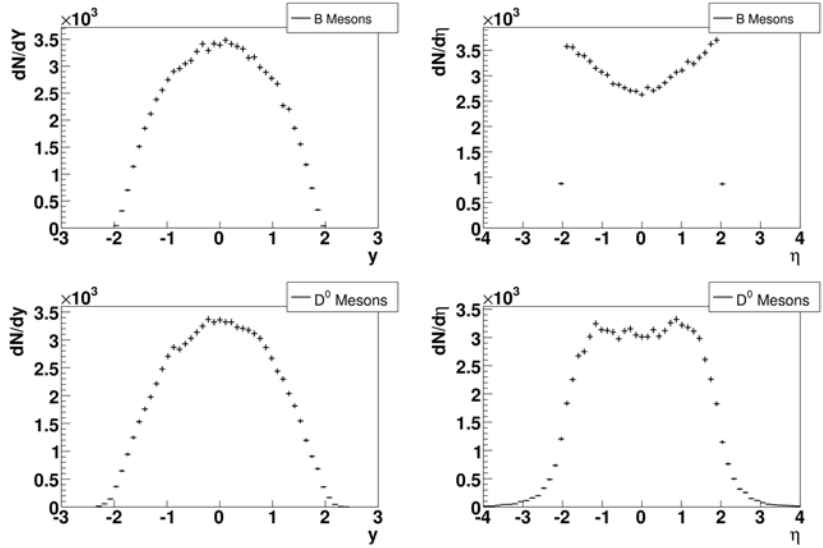

Figure 5.2: Rapidity (left) and pseudorapidity (right) distributions for B mesons and $D^0$ from B integrated in a transverse momentum for 0 to 30 GeV/c.

In Fig. 5.3 is shown the fraction of B mesons producing $D^0$ mesons with $|y| < 1$ (open symbols) and $|y| > 1$ (solid symbols) as a function of the B meson rapidity. It is interesting to note that starting with $|y| > 1.7$ all B mesons produce $D^0$s outside the acceptance of ALICE. By integrating in the range $[-1.7, 1.7]$ one can obtain the total number of $D^0$ inside rapidity $\pm 1$. Considering the branching ratios shown in Fig. 5.4 and the B meson yields from Table 5.1 we conclude that the total contribution of $D^0$ from B decays is $\sim 5.3\%$.

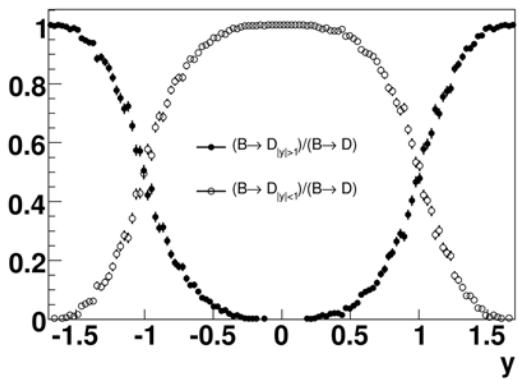

Figure 5.3: Fraction of B mesons decaying into a $D^0$

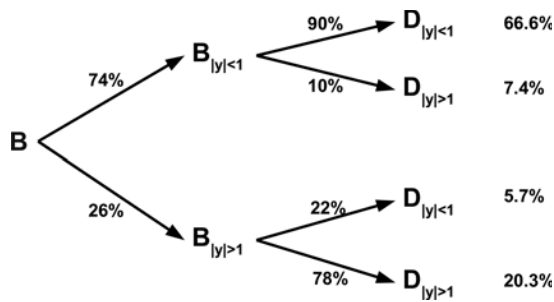

Figure 5.4: Schematic view of the gain and loss process for B mesons decaying to $D^0$. The total loss of $D^0$ around 1.7% given the rapidity distribution in Fig. 5.2.

## 5.2 Cuts

The main difference between the decay of a prompt $D^0$ and a $D^0$ originating form a B meson is the position of the secondary vertex with respect to the primary vertex. The mean decay length of the B meson is between 460 $\mu m$ ($B^0$) and 500 $\mu m$ ($B^\pm$) while for the $D^0$ it is around 123 $\mu m$. In the decay process of the B meson the $D^0$ takes most of the momentum, thus, following closely the initial direction of the B.

As a consequence the decay vertex for the $D^0$ mesons originating from B will be further away form the primary vertex than in the case of prompt $D^0$ mesons. Prolonging the reconstructed tracks of the kaon and pion back to the primary vertex will result in larger impact parameters (in absolute value) and opposite signs. In terms of the product of impact parameters this translates to a large tail on the negative side of the distribution.

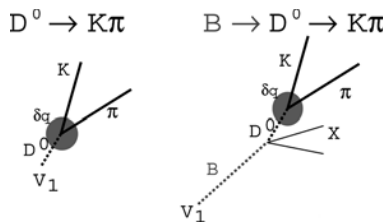

Figure 5.5: Decay topology of primary and secondary $D^0$ mesons.

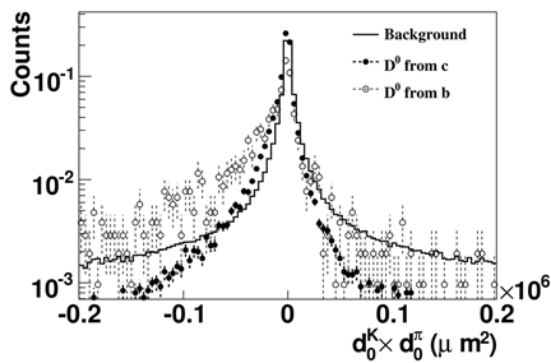

Figure 5.6: Product of impact parameters for D from beauty (open symbols), D from charm (solid symbols) and background combinations (solid line). Histograms are normalized to the same integral.

For decays where the angle between the $D^0$ and the B meson is large, the impact parameters will have both the same sign, contributing therefore on the positive side of the distribution. Figure 5.6 shows that the distribution for the product of impact parameters for $D^0$ originating from B decays has larger tails than for primary $D^0$ mesons. By placing a cut on the product of impact parameters on the negative side of the distribution ($d_0^K \times d_0^\pi < -20000 \ \mu m^2$) the fraction of reconstructed $D^0$ coming from B is enhanced with respect to primary $D^0$ mesons.

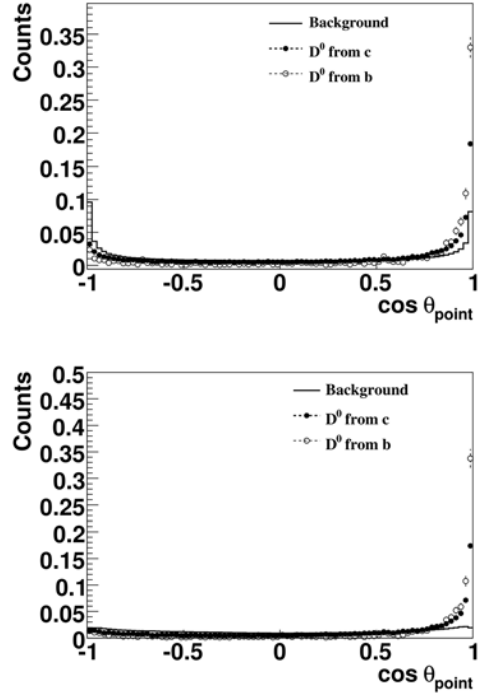

Figure 5.7: Cosine of pointing angle for primary $D^0$ (solid symbols), $D^0$ from B decays (empty symbols) and background combinations (full line). Bottom panel: a cut of $p_T^{K\pi} > 0.5$ GeV/c is applied. Histograms are normalized to the same integral.

The distribution of the cosine of the pointing angle for $D^0$ from B decays is similar to the distribution from primary $D^0$ except that the extra angle between the $D^0$ and B flight-line slightly broadens the peak at 1. Figure 5.7 shows the distribution of $\cos\theta_p$ for primary (solid symbols) and secondary (empty symbols) $D^0$ together with the background combinations. The cut on the transverse momentum of the kaon and pion ($p_T^{K\pi} > 0.5$ GeV/c) decreases the peak of the background at $\cos\theta_p = 1$. By placing a cut on the pointing angle ($\cos\theta_p > 0.9$), the fraction of primary $D^0$ particles is enhanced.

The net effect of all quality selection criteria, single and two-tracks cuts applied for the reconstructed $D^0$ mesons is that the fraction of secondaries is enhanced to 8%. Figure 5.8 shows the ratio ($D^0$ from b)/($D^0$ from c) after reconstruction (solid line) and the same ratio after cuts. Because of the harder spectrum of the $D^0$ from B mesons, the ratio increases as a function of the transverse momentum. The same trend is present also after the cuts, with a slightly larger increase in the region of low momenta. Table 5.2 summarizes the ratio as a function of the transverse momentum.

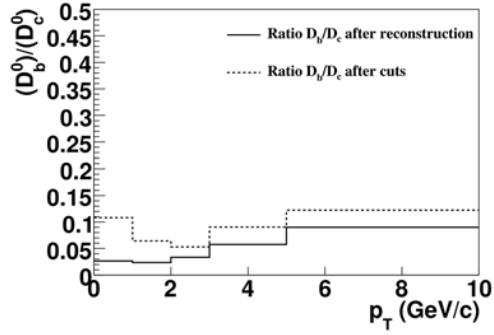

Figure 5.8: Product of impact parameters for D from b (open symbols), D from c (solid symbols) and background combinations (solid line)

| $p_T$ bin [GeV/c] | $D_b/D_c$ Before Cuts | After cuts |
|---|---|---|
| (0,1) | 0.026 | 0.108 |
| (1,2) | 0.023 | 0.064 |
| (2,3) | 0.033 | 0.053 |
| (3,5) | 0.057 | 0.090 |
| (5,10) | 0.089 | 0.122 |
| (0,10) | 0.082 | 0.034 |

Table 5.2: Total yield, average rapidity density for $|y| < 1$ and relative abundance for hadrons with charm in pp collisions at $\sqrt{s} = 14$ TeV. The contribution of $D^{*+}$ decays to $D^0$ has been taken into account.

We conclude that, with the current set of $D^0$ selections, the contamination of D mesons from B decays is increased by a factor of two. Additional studies are needed in order to understand how $D^0$ from charm can be disentangled from $D^0$ originating from beauty decays. However, since we are only interested in estimating the contamination from B decays, these studies extend beyond the scope of this thesis.

# Conclusions and Outlook

In this thesis was presented an analysis strategy for the reconstruction of open charm mesons with ALICE via $D^0 \to K^- + \pi^+$ and $D^{*+} \to D^0 + \pi^+$. Heavy flavour quarks, produced at the Large Hadron Collider in unprecedented amounts, are promising tools to investigate the properties of the hot and dense QCD medium produced in heavy-ion collisions. Baseline measurements of heavy flavour production mechanisms in pp collisions are necessary in order to understand the production in the medium.

In Chapter 3 it was shown the for one year of data taking ($\sim 10^9$ events) we will be able to measure charmed mesons with very large significances (78 for $D^0$ and 33 for $D^{*+}$). Both yields allow a measurement in one month of data taking ($\sim 10^8$ events) with a $D^{*+}$ significance of 10.

The reconstruction of the charged mesons is more difficult than for the neutral mesons due to kinematic constraints on the soft pion in case of the $D^{*+}$ decay. Most of these pions have momenta lower than 0.4 GeV/c which makes them difficult to detect and therefore the reconstruction efficiency is very low. However, by using the ALICE Inner Tracking System as a standalone detector, the reconstruction efficiencies for kaons and pions at very low momenta are improved. Kaon reconstruction efficiency improved by 11% and soft pion detection by 16%. The efficiency increase for the $D^{*+}$ with the ITS-standalone procedure after selection cuts is found to be ~10%.

For accurate measurements of charm quarks it is important to determine the contamination of the signal with $D^0$ and $D^{*+}$ from beauty decays.

Having a high purity for the D meson yields will allow a comparison between the measured $p_T$ spectrum and different theoretical predictions. It is important to note that after applying the cuts on the D mesons, the contamination with $D^0$ from beauty decays becomes slightly larger (~10%). Further analysis techniques have to be developed in order to distinguish between the two sources of $D^0$ mesons. For analyses where high efficiency is preferred over high purity, the cuts on the $D^{*+}$ and $D^0$ can be loosened in order to have as many particles as possible. In this case, a review of the contamination from B decays is needed.

# Samenvatting

In dit proefschrift wordt een analyse beschreven voor de mogelijkheid om mesonen met open charm waar te nemen met de ALICE detector via de vervalkanalen $D^0 \rightarrow K^- + \pi^+$ en $D^{*+} \rightarrow D^0 + \pi^+$.

Deze zware quarks die in grote getallen worden geproduceerd met de nieuwe LHC versneller, zijn zeer gevoelig voor de eigenschappen van het hete en dichte QCD medium. Het bepalen van zware quark productie in $pp$ botsingen is essentieel voor het begrijpen van de processen in het medium.

In hoofdstuk 3 is aangetoond dat binnen een jaar van gegevens verzamelen de charm mesonen met hoge significantie gemeten kunnen worden. Het reconstrueren van geladen mesonen is moeilijker dan van neutrale mesonen vanwege de lage transversale impuls van het extra uitgezonden pion. Deze pionen hebben meestal een impuls die lager is dan 0.4 GeV/c waardoor deze moeilijker te detecteren zijn. Echter door de ALICE ITS als 'stand-alone' detector te gebruiken kan de kaon detectie met 11% en pionen detectie met 16% verbeterd worden. Tengevolge van extra selectie criteria verbetert de totale detectie efficiency voor de $D^{*+}$ in deze configuratie met $\sim$10%.

Om precies het direct geproduceerde aantal charm quarks te bepalen moet de contaminatie via het beauty verval ook vastgelegd worden.

Indien D meson productie met grote nauwkeurigheid gemeten kan worden, is het mogelijk deze te vergelijken met de verschillende theoretische voorspellingen. Belangrijk is dat als er extra selectie criteria aan het D meson worden opgelegd, de verontreiniging via het beauty verval iets groter wordt ($\sim$ 10%). Verder onderzoek is nodig om in meer detail de twee bronnen van $D^0$ meson productie vast te leggen. Indien in een analyse een hoge efficiency belangrijker is dan zeer grote zuiverheid zouden de selectie criteria voor de $D^0$ en $D^{*+}$ minder sterk genomen kunnen worden, echter dan zal ook weer naar de bijdrage van B verval gekeken moeten worden.

# Appendix A

# Acronyms

**ALEPH** Apparatus for LEP Physics at CERN

**ALICE** A Large Ion Collider Experiment

**AliESD** Alice Event Summary Data

**CAF** CERN Analysis Facility

**CDF** Collider Detector at Fermilab

**CERN** Conseil Européen pour la Recherce Nucléaire

**CGC** Color Glass Condensate

**CINT** C INTerpreter

**FMD** Forward Multiplicity Detector

**FONLL** Fixed-Order Next-to-Leading Logarithm

**HIJING** Heavy Ion Jet Interaction Generator

**HMPID** High Momentum Particle IDentification

**ITS** Inner Tracking System

**LHC** Large Hadron Collider

**NLL** Next-to-Leading Logarithm

**NLO** Next-to-Leading Order

**PDF** Parton Distribution Function

**PHENIX** Physics Experiment at RHIC

**PHOS** PHOton Spectrometer

**PMD** Photon Multiplicity Detector

**pQCD** perturbative Quantum ChromoDynamics

**QCD** Quantum ChromoDynamics

**QED** Quantum ElectroDynamics

**QGP** Quark Gluon Plasma

**RHIC** Relativistic Heavy Ion Collider

**SDD** Silicon Drift Detector

**SLAC** Stanford Linear Accelerator

**SM** Standard Model

**SPD** Silicon Pixel Detector

**SPS** Super Proton Syncrotron

**SSD** Silicon Strip Detector

**STAR** Solenoidal Tracker At RHIC

**TOF** Time Of Flight

**TPC** Time Projection Chamber

**TRD** Transition Radiation Detector

**ZDC** Zero Degree Calorimeter

# Acknowledgements

In the beginning was the Promotor, and the Promotor was with a thesis Subject, and the Subject was Good. After a long period of time the work was done and the thesis was completed...

First, I would like to thank my promotor Rene Kamermans, my co-promotor Paul Kuijer and our group leader Thomas Peitzmann for accepting me as their student just a few days after we had the interview on that friday afternoon in 2005. I also want to thank the members of the reading committee who approved my PhD thesis four years later and allowed me to defend it in such a short period of time.

Thanks to Rene for his kind and understanding nature but also for kicking me forward when needed and make me finish the thesis on time. I would like to thank Paul for having the patience for sitting with me in front of the computer almost all day long and helping me out with Root, AliRoot and the whole physics I was looking at and had no clue what I was supposed to do. Too bad it was for less than one year. Thanks to Thomas for the many times he signed my travel papers to Geneva three or four months after I was back from the AliceWeek. Thank you also for the very nice physics discussions we had during the lectures last year when I still had time to spend on other things than working.

After Paul went to CERN and left me there with enough knowledge to be able to find my way around, I bumped my head against other kind of problems concerning my analysis. Luckily along came Marco van Leeuwen, my second co-promotor, also known as Mr. KnowItAll. Whatever question I had I could always go to his office where he greeted me with a smile on his face and solved my problem in a matter of minutes. Sometimes he would smile and make fun of me before saying anything. The only result was that I could not get mad on him but instead started thinking on my own. For some reason this way of doing things - thinking about them - managed to make me more independent and more critical. Thanks for your help!

There are so many people to mention! Thanks to the Utrecht group for the nice environment and Astrid who helped me out with bureaucratic things.

Thanks for the nice social get-togethers after lunch: Ermes, Marta, Marek, Raoul, Deepa. Thanks Alessandro for his cigarettes which, besides smoke, produced also some cool physics discussions and some plots for the thesis. Thanks to André for his feed-back after reading the thesis in just one day.

The Nikhef people should also be mentioned: Michiel for his lectures on bayesian statistics and C++, Raimond for his effort to teach the world (and me) about flow. Thanks to Mikolaj for being my linux guru and my Nikhef smoke-partner. Big thank also to Gabriel who's thesis I carried around with me for almost two years.

A warm thanks goes to the cool guy Ermes, who unconditionally endured cold winters, hot summers, rainy days and other extreme conditions whenever I felt like going out for a cigarette. Thanks for the many topical breaks, chats, laughs and the very expensive beer at JFK.

Thanks to Emanuele for the nice late-night discussions in Lunetten and Timea for helping me sneak in the neighbourhood.

A very personal note goes to Oana. I want to thank you for everything you have done for me. I would probably not be here today if it weren't for you and the patience you had with me. Thank you for your understanding and moral support, spoken or unspoken. There is much to be said... *'Hanc marginis exiguitas non caperet.'*

Before the ending credits I would like to say thanks to Ana! Your joyful and charming presence made my life during the last two months of writing more bearable. Thank you also for making the stars in Lunetten look more beautiful.

Last but not least I would like to thank my parents for encouragements and supporting my dreams. Thanks to my little sister for accepting my calls whenever I felt like talking to a friendly voice. Thanks to my broh' for coming to Utrecht so many times. The summer of 2006 will be remembered!